\documentclass[a4paper,fleqn,usenatbib,useAMS]{mnras}

\usepackage{graphicx}	
\usepackage{amsmath}	
\usepackage{amssymb}	
\usepackage{bm}		
\usepackage{pdflscape}	
\usepackage{float,textcomp}
\usepackage{natbib}
\usepackage{enumerate}
\usepackage{color}
\usepackage{lscape}
\usepackage{afterpage}
\usepackage{pdflscape}
\usepackage{subfigure}
\usepackage[T1]{fontenc}
\usepackage{ae,aecompl}
\usepackage{multicol}        
\usepackage{bm}		
\usepackage[T1]{fontenc}
\usepackage{ae,aecompl}

\title[The Void Galaxy Survey: Photometry, structure and identity of void galaxies]{The Void Galaxy Survey: photometry, structure and identity of void galaxies}

\author[B. Beygu et al.]{B. Beygu,$^{1,6}$\thanks{E-mail: burcu.beygu$@$nwu.ac.za}
R.F. Peletier,$^{1}$
J. M. van der Hulst,$^{1}$
T. H Jarrett,$^{2}$
K. Kreckel,$^{3}$ 
\newauthor
R. van de Weygaert,$^{1}$
J. H. van Gorkom,$^{4}$
M. A. Aragon-Calvo$^{5}$
\\
$^{1}$Kapteyn Astronomical Institute, University of Groningen, PO Box 800, 9700 AV Groningen, the Netherlands\\
$^{2}$Astronomy Department, University of Cape Town, Rondebosch 7700, Cape Town, South Africa\\
$^{3}$Max Planck Institute for Astronomy, K\"{o}nigstuhl 17, 69117 Heidelberg, Germany\\
$^{4}$Department of Astronomy, Columbia University, Mail Code 5246, 550 West 120th Street, New York, NY 10027, USA\\
$^{5}$University of California, 900 University Avenue, Riverside, CA 92521, USA\\
$^{6}$Physics and Centre for Space Research, North-West University, Potchefstroom, South Africa}
\date{Last updated 2016 January 7}

\pubyear{2015}

\begin{document}
\label{firstpage}
\pagerange{\pageref{firstpage}--\pageref{lastpage}}
\maketitle

\begin{abstract}
We analyze photometry from deep B-band images of 59 void galaxies in the Void Galaxy Survey (VGS), together with their near-infrared 3.6$\mu$m and 4.5$\mu$m Spitzer photometry. 
The VGS galaxies constitute a sample of void galaxies that were selected by a geometric-topological procedure from the SDSS DR7 data release, and which populate the deep 
interior of voids. Our void galaxies span a range of absolute B-magnitude from $\rm{M_B=-15.5}$ to $\rm{M_B=-20}$, while at the 3.6$\mu$m band their magnitudes range from 
$\rm{M_{3.6}=-18}$ to $\rm{M_{3.6}=-24}$. Their B-[3.6] colour and structural parameters indicate these are star forming galaxies. A good reflection of the old stellar population, 
the near-infrared band photometry also provide a robust estimate of the stellar mass, which for the VGS galaxies we confirm to be smaller than $3 \times 10^{10}$ M$_\odot$. In terms of the 
structural parameters and morphology, our findings align with other studies in that our VGS galaxy sample consists mostly of small late-type galaxies. Most of them are similar to Sd-Sm galaxies, 
although a few are irregularly shaped galaxies. The sample even includes two early-type galaxies, one of which is an AGN. Their S\'{e}rsic indices are nearly all smaller than $n=2$ in both bands and 
they also have small half-light radii. In all, we conclude that the principal impact of the void environment on the galaxies populating them mostly concerns their low stellar mass and small size.

\end{abstract}
\begin{keywords}
galaxies: evolution --- galaxies: formation ---galaxies: structure --- large-scale structure of universe
\end{keywords}

\section{Introduction}
\label{sec:intro}

Voids are prominent features of the cosmic web \citep{weyplaten2011}. Formed from 
primordial underdensities they now occupy a major fraction of the volume of the universe, 
surrounded by denser filaments, walls and sheets. They are the most underdense regions where galaxy evolution will have progressed slowly, without 
the dominant and complex influence of the environment. Voids therefore are extremely well suited for 
assessing the role of the environment in galaxy evolution, as here the galaxies are expected not to
be affected by the complex processes that modify galaxies in high density environments. The void 
environment covers the lowest density environments found in the universe, though some voids do 
approach similar (and still low) densities as found in tenuous filaments and walls \citep{marius2014}. 

In order to get a good picture of galaxies in voids the Void Galaxy Survey (VGS) was designed, 
a multiwavelength study of 59 galaxies in geometrically defined voids \citep{stanonik2009,wey2011,kreckel2011,kreckel2012,beygu2016}. Previous papers based 
on the Void Galaxy Survey have focused on the $\rm{H\textsc{i}}$ properties of galaxies in voids 
\citep{kreckel2011,kreckel2012,beygu2013} and on the star formation properties of void galaxies \citep{beygu2016}. They found that voids contain a population of galaxies that are relatively HI rich 
of which many present evidence for ongoing gas accretion, interactions with small companions and 
filamentary alignments . 

Even though based on a wide variety of selection methods, previous studies have lead to the general contention 
that void galaxies appear to be blue and low-luminosity galaxies with stellar masses lower than the average 
galaxy - typically in the order $3 \times 10^{10} M_\odot$ - and of a late morphological type, residing in a 
more youthful state of star formation and possessing larger and less distorted supplies of gas 
\citep{szomoru1996,kuhn1997,popescu1997,karachentsev1999,grogin1999,grogin2000,rojas2004,rojas2005,croton2005,goldberg2005,hoyle2005,tikhonov2006,patiri2006a,patiri2006b,ceccar2006,wegner2008,
kreckel2012,beygu2016}. \cite{penny2015} recently reported to have found some galaxies with stellar masses $\rm{10^{10}M_{\odot}}$ $<$ $M_{*}$ $<$ $\rm{5 \times 10^{11}M_{\odot}}$ 
that are located in voids, although the identification with underdense regions similar to
those of our study is not clear.

The pristine environment of voids represents an 
ideal and pure setting for the study of environmental influences on galaxy formation and evolution. The clearest 
indication for the significance of environmental influences on galaxy properties is the tight relation 
between morphology and density \citep{oem1974,dressler1980,dressler1985}. The fraction of elliptical and lenticular 
galaxies rises steeply with increasing environmental density. This goes along with the opposite trend for 
late-type and irregular galaxies, down towards the lowest density regions. This can be 
understood by observing that the evolution of galaxies in high-density regions is strongly influenced by the 
complex interplay of a range of physical processes, mostly induced by the interaction of galaxies amongst 
themselves and the intergalactic medium. Processes such as quenching, ram pressure, strangulation                                                                                                                                                              and tidal stripping render 
galaxies gas poor, yielding reddish galaxies \citep{gunn1972,larson1980,moore1996,koop2004,gabor2010,peng2010,wetzel2012}.
In more moderate and low density regions these processes cease 
to be effective, explaining the increasing fraction of late-type and gas-rich galaxies. An additional and related 
environmental influence that manifests itself in voids and that still needs to be understood is the 
finding that void galaxies appear bluer, a trend that continues down into the most rarefied void regions.

For the purpose of better understanding environmental influences on the evolution of galaxies, recent years saw a considerable increase of interest in the nature of void galaxies
\citep{kreckel2011,kreckel2012,hoyle2012,beygu2013,alp2014,moorman2014,kreckel2015,tava2015,penny2015,moorman2015,beygu2016}. Amongst the issues relevant for our understanding of 
galaxy and structure formation, void galaxies have posed
several interesting riddles and questions. Arguably the most prominent issue is that of the near absence of 
low-luminosity galaxies in voids, while standard LCDM cosmology expects voids to be teeming with dwarfs and 
low-surface-brightness galaxies \citep{peebles2001}. An interesting point of focus for void galaxy studies has 
therefore been the study of dwarf galaxies in nearby voids, such as in the Bo\"otes, Lynx-Cancer, Hercules and 
Eridanus void \citep{grogin2000, cruzen2002, pet2005, pustilnik2011a,pustilnik2013}. 
\cite{kreckel2011b} made a detailed HI study of the dwarf KK246 in the Local Supercluster, one of the darkest 
galaxies known with an M/L = 89. Another recent example concerns the study by \cite{kara2013}, 
who looked for faint void galaxies, not brighter than the Magellanic Clouds, in the Local Supercluster and 
immediate vicinity out to a distance of 40 Mpc. They found no less than 89 voids which do not appear to 
contain any galaxies brighter than $M_K<-18.4$. One of these voids is the Local Void.

Another issue of interest is whether we can observe the intricate filigree of substructure in voids, expected as 
the remaining debris of the merging of voids and filaments in the hierarchical formation process \citep{weykamp1993,sheth2004,weyplaten2011,aragon2012}.
Evidence for such substructure, three interacting galaxies embedded in a 
common HI envelope, has been reported by \cite{beygu2013}, who hypothesised it to be an assembly of 
a filament in a void.

Of key importance towards deciphering the nature and evolutionary history of void galaxies are their structural 
properties. Most of the previous observational studies of void galaxies were based on an analysis of photometric 
data from existing all sky surveys, such as the Sloan Digital Sky Survey and the Center for Astrophysics  
Redshift Survey \citep{hoyle2002a,rojas2004,rojas2005,patiri2006b,hoyle2012,tava2015,alp2015,penny2015}. Lately, also within the context of the Galaxy Mass and Assembly Survey 
(GAMA, \cite{driver2011}) considerable attention has been devoted to voids and void galaxies \citep{alp2014,alp2015,penny2015}. Using catalogues of large-scale structure including 
voids, group and pair membership from the GAMA survey, \cite{alp2015} examined the galaxy properties and found that the stellar mass is the dominant factor in 
shaping the galaxy properties. Such studies  are limited in depth 
or resolution because of the magnitude limits of the surveys. The present study of the galaxies in the 
Void Galaxy Survey focusses on the analysis of deeper photometric data that we obtained for the VGS galaxies, to 
assess their colour, stellar mass, galaxy concentration, morphology and specific star formation. 
We use deep B-band and 3.6$\mu$m near-infrared imaging to investigate the structural characteristics and morphologies of galaxies 
in the Void Galaxy Survey (VGS, see section 2). The analysis involves the determination of the total luminosity, 
characteristic scale $\rm{r_{e}}$ and the surface brightness $\rm{\mu_{e}}$ of the galaxies, along with the concentration of the 
stellar population, quantified by the Sersic index $n$. The inferred measurements are compared to literature values 
of structural parameters for a wide range of different galaxies. Amongst these are late-type disk galaxies, dEs as 
well as giant early-type galaxies. 

By using photometry in three bands, two near-infrared and one optical, our photometric study yields important
insights into the morphology and stellar populations of the VGS galaxies. The deep B-band images trace both the 
younger and older population. A unique aspect of our study is the inclusion of the Spitzer 3.6$\mu$m and 4.5$\mu$m imaging data. In particular the 3.6$\mu$m band data represents a major asset 
for our understanding of the stellar population of these galaxies, given the insensitivity of the 3.6$\mu$m band images to dust extinction and the fact that they 
provide a good reflection of the old stellar population in these galaxies (see e.g. \cite{peletier2012} and 
\cite{meidt2014} for a discussion). This also means that the 3.6$\mu$m flux, together with the [3.6]- [4.5] colour provide us with a robust estimate 
of the stellar mass of a galaxy, since the old stellar population constitutes its major share. In addition, by 
combining information on the young and older stellar populations, the B-[3.6] colour 
provides us with a good indicator of the composition of the stellar population.
 
In addition, we seek to extract information on the star formation and evolution of void galaxies.
From the colour comparison of B-band photometry with the Spitzer 3.6$\mu$m band photometry of the void galaxies, 
we assess their star formation histories. This information is combined with the results obtained from 
near-UV imaging to infer colours and the specific star formation rates $SFR_{NUV}/M_{*}$ 
(see \cite{beygu2016}). 

The paper is organized as follows: In section~\ref{sec:vgssample} we describe the VGS sample and the results found so far. 
Section~\ref{sec:observ} contains a description of the observations and the data analysis. In section~\ref{sec:morphology_structure} we subsequently 
present the morphology of the void galaxies, and attempt to relate this to their underdense void environment. 
In the same section, we present and briefly discuss the structural parameters of these galaxies. The star formation 
properties and evolution of the stellar population form the subject of section~\ref{sec:stel_pop}. Finally, in section~\ref{sec:discussion} and 
section~\ref{sec:conclusion} we shortly discuss and summarize our findings.

\section{The VGS sample}
 \label{sec:vgssample}
The Void Galaxy Survey (VGS) \citep{wey2011,kreckel2011,
 kreckel2012,beygu2016} is a systematic multiwavelength survey of 59 void galaxies, aiming to probe the colour, morphology, star 
 formation and gas content of the void galaxies. For this we observed the VGS galaxies at radio wavelengths, in the 21-cm line, in 
 $\rm{H \alpha}$ and in the optical B-band. In addition, we acquired GALEX near-UV data on the VGS galaxies, as well as Spitzer 
 3.6$ \rm{\mu m}$ imaging. 
 For some we obtained CO(1-0) observations, which will form the starting point for a study of the relation between their 
 star formation activity and their molecular and atomic gas content. 

 Our VGS galaxy sample has been selected from the Sloan Digital Sky Survey Data Release 7 (SDSS DR7). We exclusively used  
 geometric and topological techniques for delineating voids in the galaxy distribution and identifying galaxies populating 
 the central interior of these voids \citep{schaap2000,platen2007,aragon2010a,kreckel2011}. The typical size of voids in our sample is 
 on the order of 5 to 10 Mpc. The resulting sample of void galaxies is unbiased and largely independent of intrinsic galaxy properties 
 (except for the selection criteria defined by the SDSS). The importance of this is that it may be used towards obtaining an optimally 
 representative census of the properties of galaxies that are located in the most diluted regions of the Universe. 
  The detection of the void galaxies is limited to the spectroscopic flux limit of 17.7 mag in the r-filter of the SDSS. 
 The VGS galaxies have redshifts in the range $\rm{0.02 <  z < 0.03}$, 
 while they have an absolute magnitude in the range of $\rm{-20.4 < M_{r} < -16.1}$, and colours in between $\rm{0.6 < g-r < 0.87}$ \citep{kreckel2011,kreckel2012}.
 
 So far we have completed the study of the $\rm{H\textsc{i}}$ \citep{kreckel2012} and star formation properties \citep{beygu2016} of the VGS sample. We detected 41 of them in 
 $\rm{H\textsc{i}}$ , with total masses ranging from  1.7 $\times$ $\rm{10^8}$ to 5.5 $\times$ $\rm{10^{9}}$ $\rm{ M_{\odot}}$. We found 
 that many have extended $\rm{H\textsc{i}}$ disks. Often these are morphologically and kinematically disturbed, showing signs of ongoing 
 gas accretion. In addition, we found 18 $\rm{H\textsc{i}}$-rich neighbouring galaxies within a radial distance of 100 km $\rm{s^{-1}}$ from 
 the targeted galaxy. However, we have not been able to establish the presence of any significant number of $\rm{H\textsc{i}}$-rich 
 low-luminosity galaxies that would fill the void \citep{kreckel2012}. As such, we found no indication for the existence of a large population of 
 dwarf galaxies, whose presence would be expected according to theoretical expectations \citep{peebles2001}.
 
 What we did find are 
 several interesting and at first unexpected peculiar void galaxies. One of these is a polar ring galaxy, VGS\_12, located in a tenuous 
 wall between two voids \citep{stanonik2009}. We also found a configuration of three void galaxies, VGS\_31, aligned along what appears to be a tenuous filament along a 
 void \citep{beygu2013,rieder2013}. We compared their $\rm{H \alpha}$ star formation rates per stellar and per
$\rm{H\textsc{i}}$ mass with those for galaxies in moderate underdensity regions of the cosmic web and found only a marginal increase in star formation rates \citep{beygu2016} Their 
optical emission line and infrared colour properties show that the VGS galaxies are star forming galaxies with the exception of one passive galaxy.
\section{Observations}
\label{sec:observ}
B-band imaging has been obtained with the Wide Field Camera (WFC) at the 2.5m Isaac Newton Telescope (INT) at the Observatorio del Roque de los Muchachos at La Palma using 
the Johnson-Cousins B-filter between March 2010 and March 2012. Total exposure times 
were 2400 seconds, spread over 4 exposures for the purpose of dithering and facilitating cosmic ray detection. The average seeing was 1.4 arcsecond and the pixel scale is 0.33 arcsec/pixels.
Flat-field exposures were taken at twilight at the beginning and/or end of each night. 

The near-infrared data have been obtained from the Spitzer 3.6$\rm{\mu m}$ and 4.5$\rm{\mu m}$, observed for the program 80069. We used the
 Spitzer-IRAC [3.6] Level 2 (post-BCD) data, which have been calibrated by the Spitzer-IRAC pipeline. The pixel scale is 0.6 arcsec/pixels. To understand the IRAC 3.6 micron PSF we 
note that the original pixel size was 1.2'', and that the data were severely undersampled (the diffraction limit of Spitzer at 3.6 micron is only 1.05'' (FWHM)), giving an effective 
 $\rm{FWHM_{PSF}}$ of about 1.6''. Here we do not use any data inside 2.4'' for fitting the surface brightness distributions. 
\begin{figure}
 \centering\includegraphics[scale=0.2]{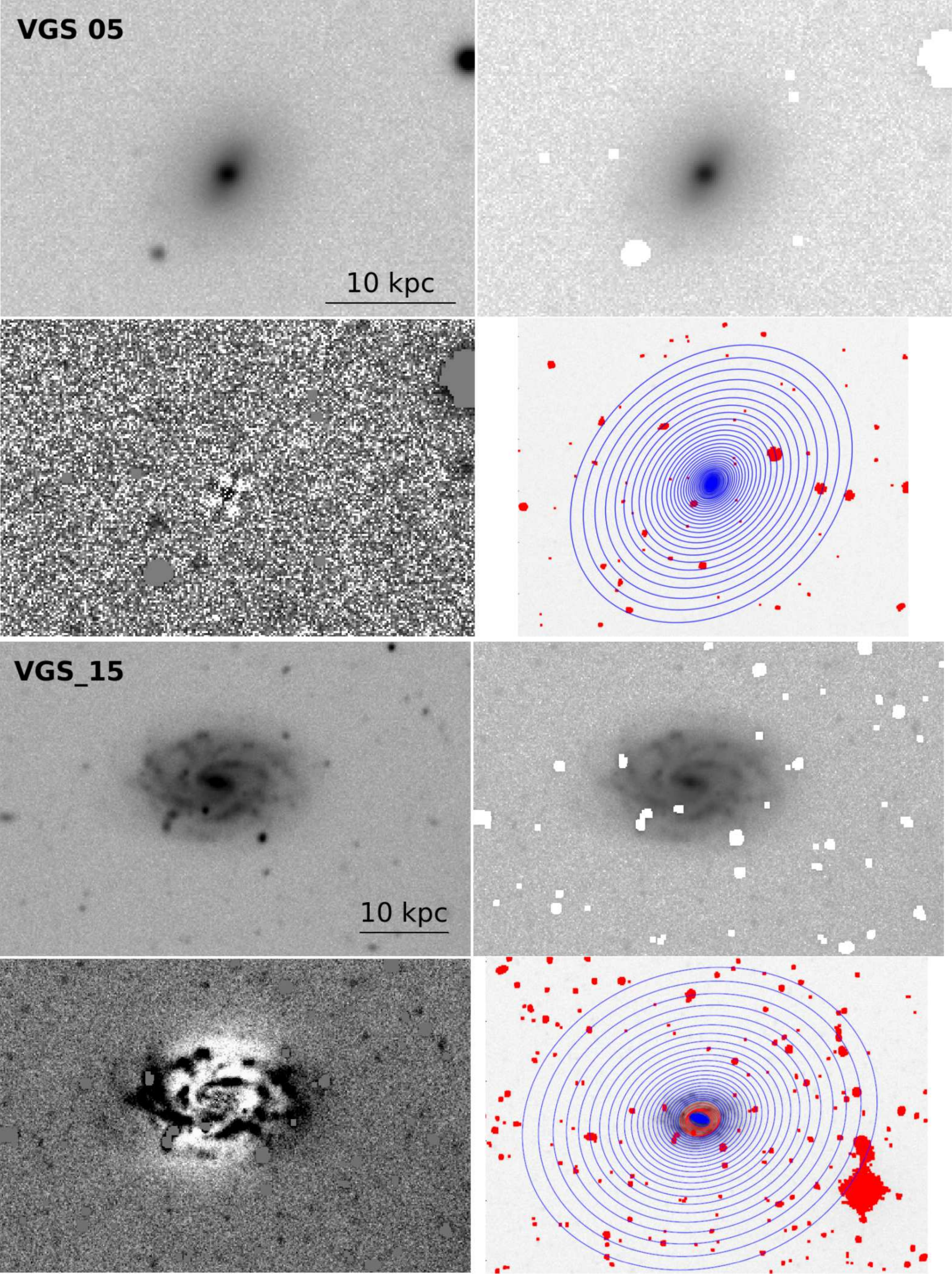}
  \caption{Raw, cleaned and model-subtracted images and resulting ellipse fits for B-band created by ARCHANGEL (clockwise from left) for VGS\_05 (top) and VGS\_15 (bottom). 
  (A colour version of this figure is available in the online journal.) } 
 \label{fig:arch_B}
 \end{figure}
\subsection{Data reduction}
\label{sec:data reduction}
The INT data have been reduced using the standard IRAF\footnote{http://iraf.noao.edu/} procedures for CCD imaging. All the optical images were
trimmed and overscanned, followed by bias subtraction and flat fielding. After that all images from each filter were aligned and median combined.
Photometric calibrations have been done by calibrating each science frame using 3 SDSS stars with g $<$ 17. 
To do this, SDSS g band magnitudes were converted to the Johnson-Cousins B photometric system following the relation given in \cite{fuk1996}:

\begin{equation}
 g-r    =    0.93 \times (B-V)   - 0.06,
\end{equation} 

We converted our IRAC counts to magnitudes on the Vega magnitude system, using the fact that IRAC 3.6$\rm{\mu m}$ has a flux density of zero magnitude source
 of $\rm{F_{\upsilon}}$= 280.9 Jy, and 4.5$\rm{\mu m}$ has $\rm{F_{\upsilon}}$= 179.7 Jy, 
implying that 1 count/sec corresponds to 18.80 Mag in the [3.6] band, and 18.32 in the [4.5] (IRAC Instrument Handbook\footnote{http://irsa.ipac.caltech.edu/data/SPITZER/docs/irac/}).
\begin{figure*}
 \centering\includegraphics[scale=0.35]{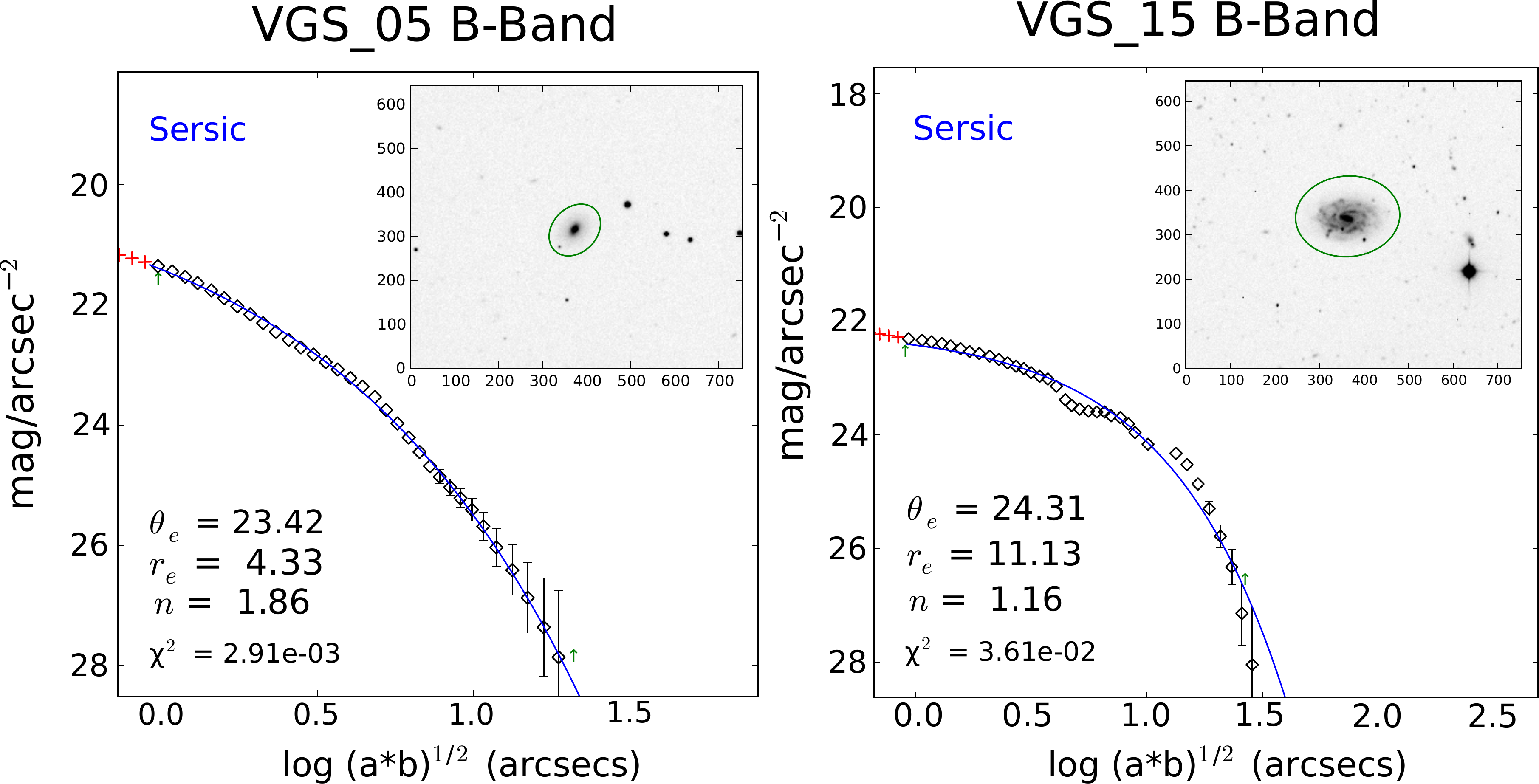}
  \caption{Final surface photometry profile for VGS\_05 and VGS\_15. Solid lines are the fit to a one-component S\'{e}rsic function. a and b are the major and the minor axis of the 
ellipse, respectively. Red crosses represent the inner cut-off regions. (A colour version of this figure is available in the online journal.) } 
 \label{fig:sersic}
 \end{figure*}

After this, aperture photometry has been performed using the galaxy photometry package ARCHANGEL \citep{sch2007} for both B-band IRAC [3.6] and [4.5] band images.
The Archangel photometry tool performs the following steps: i) cleaning, flattening and sky determination of the science frame. Foreground stars are masked during the
cleaning procedure. The sky is determined using sky boxes placed in the frame avoiding stars and far away from other galaxies.
An iterative mean and $\rm{\sigma}$
are calculated for each box. They are then averaged to find the value of the sky. 
ii) isophotal fitting where elliptical isophotes are used. Archangel allows that position angles and ellipticities can be kept frozen at all radii. In this work these parameters are 
freely determined. 
Any pixels above (or below) a multiple of the RMS around the ellipse are masked along an isophote. These  masked regions are later restored for
aperture photometry. iii) conversion of the 2D information to 1D surface brightness profiles. iv) profile fitting. 

Figure~\ref{fig:arch_B} shows raw, cleaned, model-subtracted and ellipse-fitted images of two galaxies (VGS\_05 and VGS\_15) as examples. An inner radial cutoff has been applied 
to account for the uncertainty caused by the seeing for
the B-band images. Upper radial cutoffs are put where the background starts to dominate. In the case of Spitzer images, the inner cutoff is always constant and equals 2.4 arcseconds. 
In the B-band it 
is twice the FWHM of the seeing, after \cite{franx1989} and \cite{peletier1990}.
Following this, total magnitudes ($\rm{m_{tot}}$) and effective radii were derived using curves of growth from the ellipse-fitting results for both wavelengths. Given it is not model dependent, 
this produces reliable results. For the Spitzer IRAC imaging errors on the total magnitudes are dominated by the sky background errors which are $~$0.1 mag. As for the 
 B-band, errors on the total magnitudes are smaller than a combination of the errors on the zeropoint and the sky background, which is $~$0.2 mag. 
 
The B-band magnitudes
are corrected for galactic extinction using \cite{sch1998}. The effective surface brightness, $\rm{\mu_{e}}$, is defined as the surface brightness at the effective radius
$\rm{r_{e}}$ (half-light radius) and is determined using curves of growth. It is defined as the radius at which the magnitude reaches: 

\begin{equation}
m(r_{e}) = m_{tot}+2.5 \times log2,
\end{equation}

Surface brightness profiles have been fit to a one component S\'{e}rsic function \citep{graham2005} that gives the S\'{e}rsic index \textit{n}, such that

\begin{equation}
 I(R) = I_{e} \, \exp \Biggl\{ -b_{n} \left[ \bigl( \frac{R}{R_{e}} \bigr) ^{1/n} - 1 \right ] \Biggr\},
\end{equation}
where $n=1$ is typically assumed to correspond to the exponential disk profile of disk galaxies, and $n=4$ to the de Vaucouleurs profile of ellipticals. 
In Figure~\ref{fig:sersic}, S\'{e}rsic profile fitting
for VGS\_05 and VGS\_15 is shown for the B-band.

Traditionally, late-type galaxies are described in terms of their disc scalelength, $h$, and central surface brightness, $\rm{\mu_{o}}$. The disk and bulge are often fit separately as 
well. In our study, we are challenged by the relatively low resolution of the data (caused by seeing and sampling) compared to the size of the galaxies, so we only fit one S\'{e}rsic law,
a technique which is often used for distant galaxies. We have also applied a pure disk profile using a bulge + disk fit with a zero-sized bulge according to \cite{free1970} and 
\cite{kent1985} as:
 
\begin{align}
 I(R) = I_{0} \, \exp(-hR), \\
 \mu(R) = \mu_{0} + 1.082 \left( \frac{R}{h} \right), 
\end{align}

We have converted our effective surface brightness, $\rm{\mu_{e}}$, to the mean effective surface brightness $\rm{\langle\mu\rangle_{e}}$, assuming 
the S\'{e}rsic index $n=1$, following the recipe in \cite{graham2005}:
\begin{align}
 \langle\mu\rangle_{e} = \mu_{e} - 0.699
\end{align}
Stellar masses are calculated using the [3.6]-[4.5] colours via the empirical mass-to-light ratio given in Querejeta et al. 2014 as:
\begin{align}
log(M/L)=-0.339 (\pm 0.057) \times ([3.6]-[4.5]) -0.336 (\pm 0.002)
\end{align}
for an assumed M/L$_{3.6 \mu m}=0.6~M_\odot/L_\odot$ for the old stellar population.
\subsection{Consistency assessment of the data }
\label{sec:con_asmnt}
\begin{figure}
\centering
  \includegraphics[scale = 0.27,angle=270]{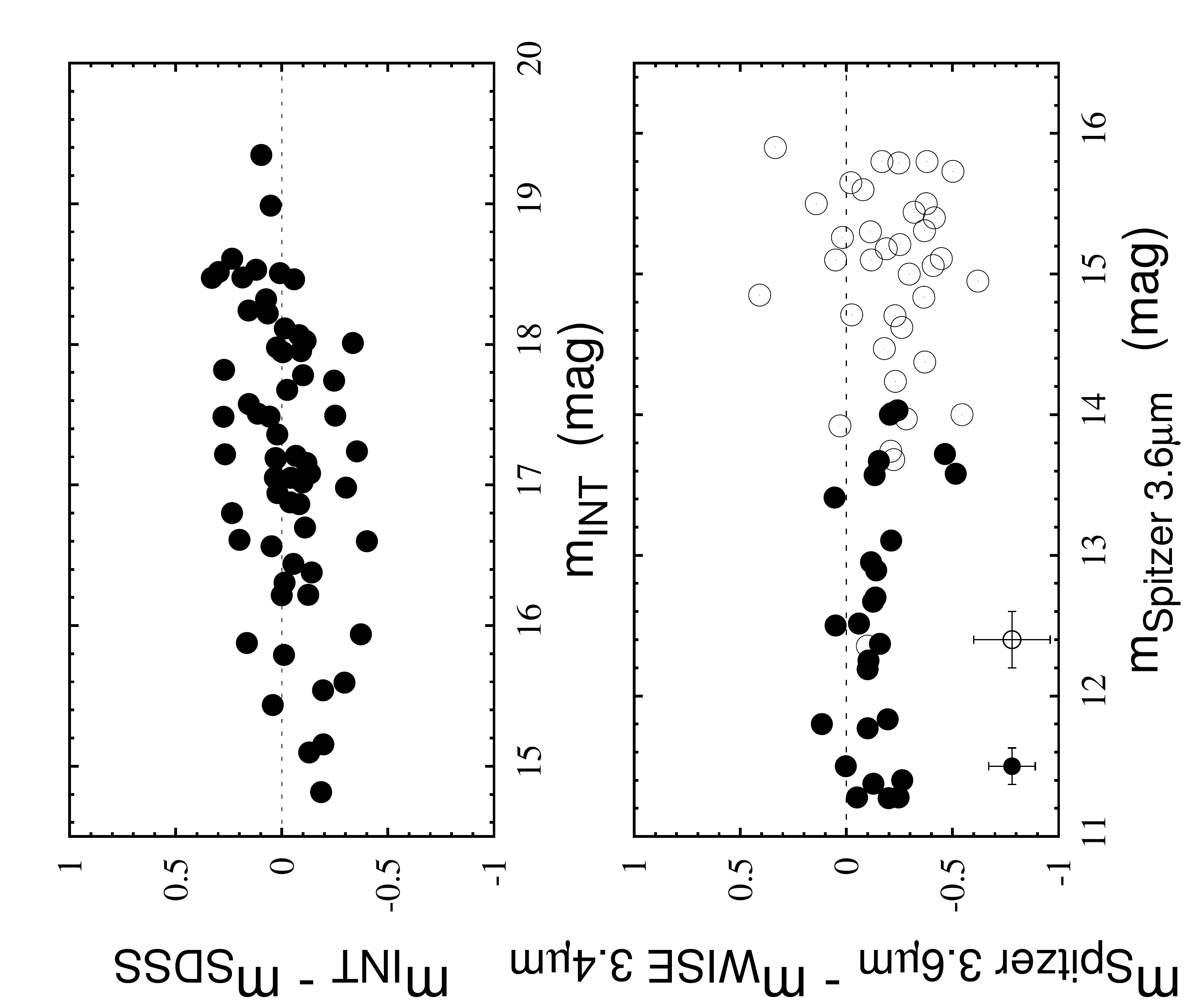}
\caption {Top: INT B magnitudes compared to SDSS g-band model magnitudes converted to the Johnson-Cousins B system. 
Bottom: Spitzer [3.6] band magnitudes compared to WISE [3.4] band
 magnitudes. Open circles are defined as point source by the WISE pipeline, for which $\rm{\chi^{2}}$ $<$ 2 for the 3.4$\rm{\mu m}$ profile-fit photometry. } 
  \label{fig:sdss_int_spit}
\end{figure}

To check the consistency of our measurements, we compare the various structural parameters 
with values found in the literature. The optical B-band total magnitude, effective radius and stellar mass estimates of the galaxies in 
our sample are compared with literature values obtained on the basis of SDSS data. We converted the SDSS g-band model magnitudes to the Johnson-Cousins B system and compared them to our 
B-band magnitudes in Figure~\ref{fig:sdss_int_spit} (top panel). The mean difference is $\sim$ -0.02 with $\rm{\sigma}$ $\sim$ 0.2 magnitude.

The effective radius is determined on the basis of a S\'{e}rsic fit. The stellar mass estimates are obtained from the MPA/JHU catalogue for SDSS DR7\footnote{The MPA-JHU catalogue is publicly 
available and may be downloaded from http://www.mpa-garching.mpg.de/SDSS/DR7/archive.}. The comparison also involves the magnitudes 
from the near-infrared Spitzer [3.6] band photometry, which is compared to  WISE [3.4] band (W1)
magnitudes, and the stellar mass inferred from the [3.6] band image.  
 
We compared our [3.6] band magnitudes to the $\rm{1\sigma}$ isophotal magnitudes of W1 which are determined with profile-fitting photometry 
(Figure~\ref{fig:sdss_int_spit}, bottom panel). The reduced chi-square, $\rm{(\chi^{2})_{W1}}$, of this isophotal fit gives a measure for an object to be classified either as a point source or a resolved 
source. If the $\rm{(\chi^{2})_{W1}}$ $<$ 2, then the object is considered as an unresolved source (point source), otherwise a resolved source (extended source). Due to their small size, 
most of the unresolved VGS galaxies are not resolved by WISE and Spitzer. Some are heavily contaminated by the foreground stars, making it 
difficult to perform aperture photometry on these objects. In the bottom panel of Figure~\ref{fig:sdss_int_spit}, we use the total magnitudes determined by isophotal and extrapolated fits for the objects 
resolved by WISE and magnitudes determined by isophotal fits for those that are unresolved. In this figure open circles represent VGS galaxies that are not resolved in W1 
and the filled circles represent those that are resolved. There is a significant difference between the residuals of the resolved and the unresolved VGS galaxies. The difference is most 
likely caused by the fact that the WISE magnitudes are isophotal magnitudes; given the much higher S/N of the Spitzer data, we argue that our measurements are fine, with scatter < 0.3 mag.

\begin{figure}
 \centering\includegraphics[scale=0.5,angle=270]{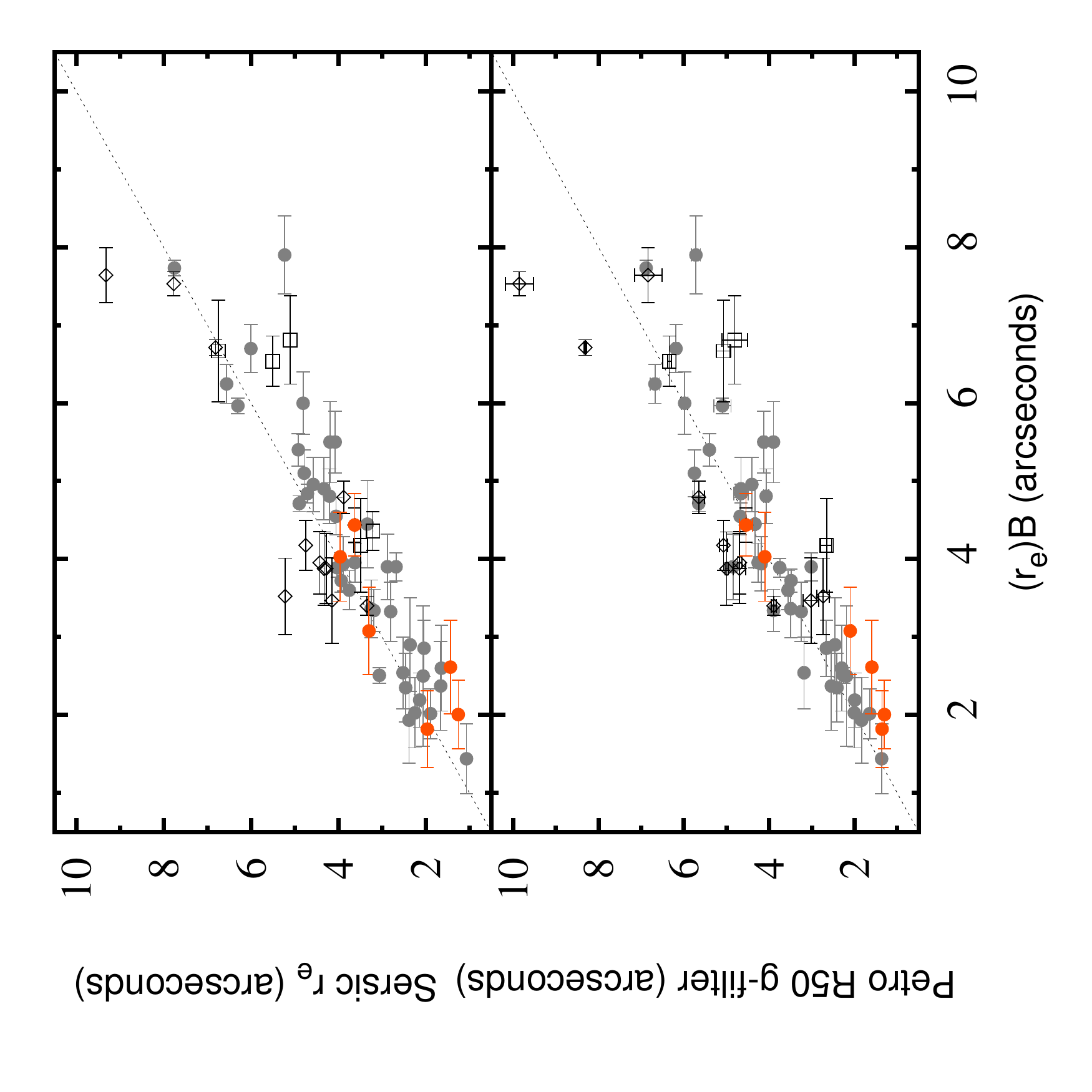}
 \caption{Comparison of $\rm{(r_{e})_{B}}$ to the S\'{e}rsic effective radius (top) and to the SDSS Petrosian half-radius R50 in g-filter (bottom). Galaxies with 
$\rm{FWHM_{PSF}}$ $<$ 2 arcseconds are shown as grey and those with $\rm{FWHM_{PSF}}$ $\geq$ 2 arcseconds are shown as orange. Open squares represent irregular galaxies and 
 open diamonds represent edge-on galaxies. See main text for the definition of the classification for ``irregular'' and ``edge-on''. The unity line is shown as dashed grey line. (A colour version of this figure is available in the online journal.)}
 \label{fig:re_comp}
 \end{figure}

In order to show the effects of the seeing, sky and the limited radial 
range on our photometric analysis we compare the effective radii in the B-band to those derived via 
S\'{e}rsic fits (S\'{e}rsic $\rm{r_{e}}$) (Figure~\ref{fig:re_comp}, top panel) and to the SDSS Petrosian half-radius R50 in the g-filter ($\rm{R_{P50}}$) (Figure~\ref{fig:re_comp}, bottom panel). Errors on the
$\rm{(r_{e})_{B}}$ are due to the uncertainties in the accuracy of sky subtraction.
The SDSS algorithm uses a modified version of the Petrosian system \citep{petrosian1976} that measures galaxy fluxes 
within a circular aperture, whose radius ($\rm{r_{P}}$) is defined by the shape of the azimuthally averaged light profile\footnote{http://www.sdss.org/DR7/algorithms/}. $\rm{R_{P50}}$ is defined as 
the radius containing 50$ \% $ of the flux within a certain number of $\rm{r_{P}}$. 

In Figure~\ref{fig:re_comp}, galaxies which have an irregular appearance are shown as open squares. Galaxies that have inclination i $>$ $70^{ \circ }$ according 
to \cite{kreckel2012} are classified as edge-on and are shown as open diamonds in Figure~\ref{fig:re_comp}. Galaxies that have relatively
poor seeing ($\rm{FWHM_{PSF}}$ $\geq$ 2 arcseconds) compared to the rest of the sample are shown in orange dots. The difference between our effective radii and the effective radii determined in
both other ways are larger for these galaxies. The mean of $\rm{(r_{e})_{B}}$-to-S\'{e}rsi$\rm{c_{r_{e}}}$ ratio is $\sim$ 1.1 and the
mean of $\rm{(r_{e})_{B}}$-to-$\rm{R_{P50}}$ ratio is $\sim$ 1.04 (VGS galaxies that are neither edge-on, irregular or that have poor seeing, are shown as grey dots in Figure~\ref{fig:re_comp}). 
We expect the agreement with the S\'{e}rsic half-light radii to be worse, since the method to determine them is much less stable than the method using apertures. Our $\rm{(r_{e})_{B}}$ 
estimates are in agreement with the Petrosian half-radius R50 and the S\'{e}rsic effective radius according to the relation shown in \cite{graham2005}. In the analysis that follows, we will use the 
effective radius determined via the curve of growth, because it is independent of any assumed model and the effective surface brightness that corresponds to this $\rm{r_{e}}$.
\begin{figure*}
 \centering\includegraphics[scale=0.12]{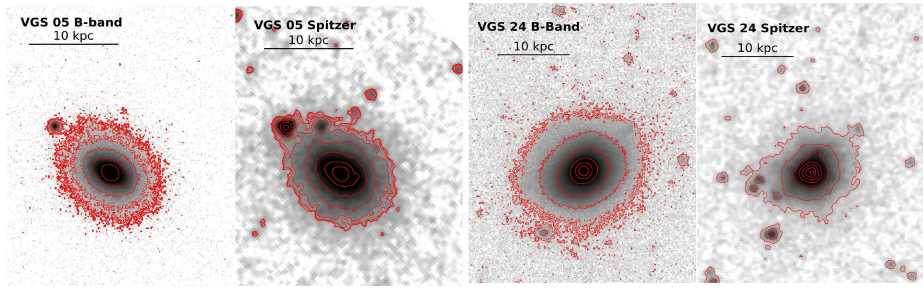}
 \caption{Examples of the different morphological representatives among the VGS void galaxies. For each morphological representation two galaxies are shown in both B and [Spitzer 3.6] band. In each image, the black bar 
 represents a physical scale of 10 kpc. (A colour version of this figure is available in the online journal.)}
 \label{fig:morph}
 \end{figure*}
\section{Morphology and structure of the VGS galaxies}
\label{sec:morphology_structure}
In this section, we will discuss the morphology and structural parameters of the VGS galaxies. By comparing magnitude, S\'{e}rsic index, effective radius and 
effective surface brightness of our sample to the same properties of other galaxies of different Hubble types,
we seek to learn more about the identity of the void galaxies. In addition, we will discuss the 
properties of the galaxies, including their $\rm{H\textsc{i}}$ content and the relationship with the low density of their void environment.  
\subsection{Morphology of the VGS galaxies} 
\label{sec:morph_den}
Morphological classification of galaxies involves rather specific issues, amongst others, high spatial resolution. An accurate classification requires a 
more elaborate analysis \citep[see][]{buta2013} than we are able to provide here. Therefore, instead of carrying out an absolute 
morphological classification, we seek to classify the morphology of the VGS galaxies in a general way by eye. 
We find that the VGS galaxy sample mainly consists of disk galaxies with an occasional bar and spiral structure. Amongst the VGS galaxies, a few objects are small and compact. 
Seven VGS galaxies have an irregular shape, twelve are edge-on systems 
(i > $\rm{70^{\circ}}$) while two objects are early-type galaxies.
Five VGS galaxies --- VGS\_30, VGS\_31, VGS\_37, VGS\_38 and VGS\_54 --- have nearby companions within 50 kpc. Some of these 
show signs of interactions, either in their 
optical or $\rm{H\textsc{i}}$ morphology, or in their kinematics \citep{kreckel2012,beygu2013}. 

Eleven galaxies have neighbours and companions between 100 to 600 kpc, detected in $\rm{H\textsc{i}}$. These galaxies do not have any 
$\rm{H\textsc{i}}$ connection with the main VGS galaxy and there is no spectral redshift information for them in the SDSS database. 
Also we do not have any information on their dynamical or stellar mass. Moreover, except for these companions our B-band images do not reveal a significant increase of dwarf galaxy resembling objects 
within 500 kpc distance from the identified void galaxy.
\begin{figure*}
\subfigure[]
{
  \includegraphics[scale = 0.42,angle=270]{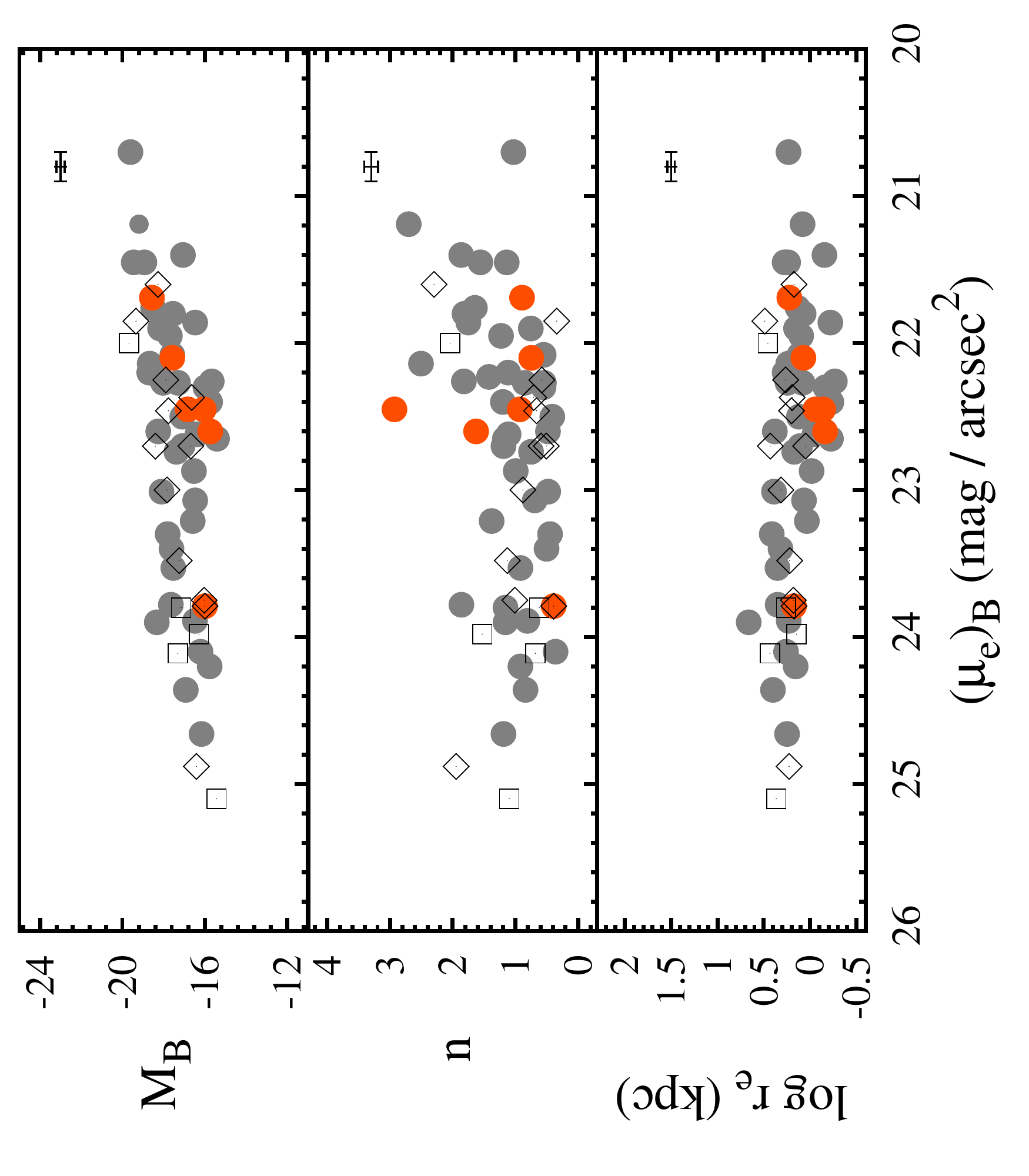}
\label{fig:first_sub}
}
\subfigure[]
{
  \includegraphics[scale = 0.42,angle=270]{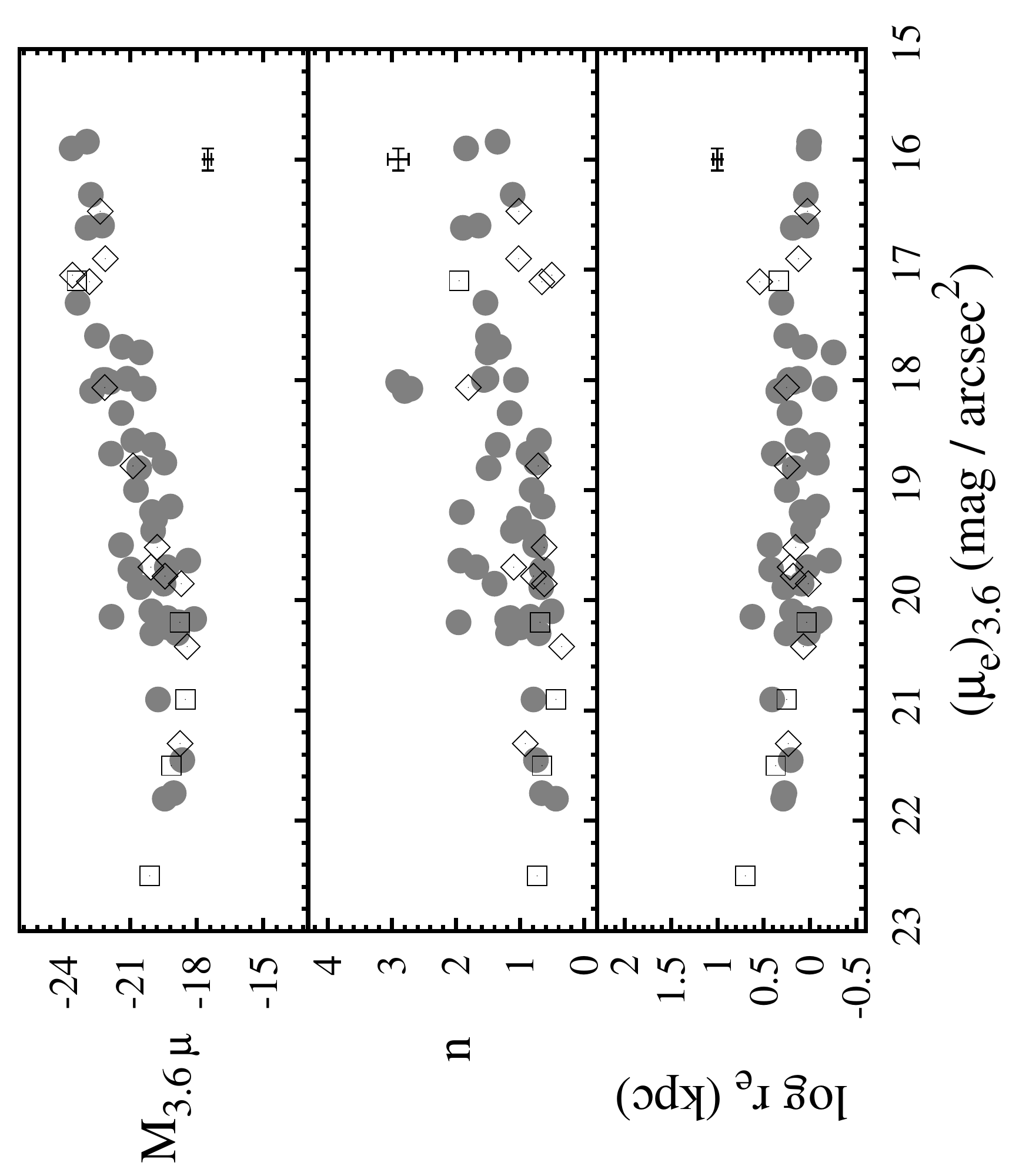}
\label{fig:second_sub}
}
\caption {Absolute magnitude, S\'{e}rsic index $n$ and the logarithm of the half-light radius $\rm{r_{e}}$ shown against the surface brightness at $\rm{r_{e}}$ $\rm{(\mu_{e})}$ for B-band (a) and Spitzer 
[3.6] band (b). The symbols are the same as in Figure~\ref{fig:re_comp}, except that in (b) we do not distinguish galaxies depending on their seeing quality in the B-band. Typical error bars are shown
in black on the righthand side of each plot. (A colour version of this figure is available in the online journal.)  } 
  \label{figure:phot}
\end{figure*}
Examples representing the different types of galaxies that constitute our sample are presented in Figure~\ref{fig:morph}, 
in both B-band and [3.6] band. One of them, VGS\_15, is a face-on spiral galaxy with tight spiral arms. These are clearly 
visible in the B-band, but not as pronounced in 3.6$\rm{\mu m}$. Its nuclear bulge is visible in both wavelengths.
VGS\_57 is a spiral galaxy with a bar that is bright in both wavelengths. VGS\_21 is an edge-on spiral, whose 
nuclear bulge is clearly seen in 3.6$\rm{\mu m}$ but obscured by dust in B-band. VGS\_50 may be an example of an edge-on 
S0 type galaxy. Its bulge is visible in both wavelengths, whereas its disk is only visible in the B-band. VGS\_17 and VGS\_38 
are examples of VGS galaxies with irregular shapes. VGS\_38 is a system of three galaxies embedded in the same 
$\rm{H\textsc{i}}$ envelope \citep{kreckel2012}. In Figure~\ref{fig:morph}, all three members of the system are shown.  
VGS\_20 and VGS\_41 represent what we call compact objects in our sample due to the lack of any distinguishable feature. 
Both of them are very blue objects without an $\rm{H\textsc{i}}$ detection. VGS\_05 and VGS\_24 are two early-type galaxies, of which VGS\_24 is the only one with an identified
AGN in our sample \citep{beygu2016}. Neither of these are detected in $\rm{H\textsc{i}}$.

In the following section~\ref{sec:Structural parameters of the VGS galaxies}, we find that also on the basis 
of their half-radii and S\'{e}rsic indices, most VGS galaxies appear to be rather small late-type disk galaxies.

\subsection{Structural parameters of the VGS galaxies}
\label{sec:Structural parameters of the VGS galaxies}
Figure~\ref{figure:phot} shows the main photometric results in the B-band (Figure~\ref{figure:phot}a) and in the [3.6] band (Figure~\ref{figure:phot}b). Each diagram consists 
of three panels. From top to bottom these plots show the absolute magnitude, S\'{e}rsic index $n$ and the half-light radius $\rm{r_{e}}$ as a function of the the surface brightness 
$\rm{\mu_{e}}$ at radius $\rm{r_{e}}$. VGS galaxies that are irregular in shape, are edge-on (i $>$ 70$^{\circ}$) and/or have $\rm{FWHM_{PSF}}$ $\geq$ 2 arcseconds (only for B-band) are
presented with the same symbols as in Figure~\ref{fig:re_comp}.  
Typical error bars are shown in black at the righthand side of each plot. Errors on the surface brightness are a combination of errors in the zero point calibration and the formal errors from the fit. Similarly, errors
in the absolute magnitudes result from  errors in the zero point calibration as explained in Section~\ref{sec:observ}. Errors in the half-light radii errors are due to the uncertainties in the sky level and they are small.
The error bars are small compared to the large parameter ranges in Figure~\ref{figure:phot}. And errors on the S\'{e}rsic indices are formal errors from the fit.
We list the photometric results of the VGS galaxies in Table~\ref{table:2}.

The majority of the VGS galaxies have S\'{e}rsic indices \textit{n} $<$ 2 in both bands and have half-light radii $<$ 3.3 
kpc, which confirms that they are small, late-type disk galaxies. A similar result is also found by \cite{alp2015} using r-band photometry. The surface brightness varies between 
25 and 20 $\rm{mag/arcsec^{2}}$ in the B-band, whereas the 3.6 $\rm{\mu m}$ 
surface brightness of the galaxies covers a wider range than in the B-band, from 23 to 15.5  $\rm{mag/arcsec^{2}}$.

\begin{figure}
 \includegraphics[scale=0.45,angle=270]{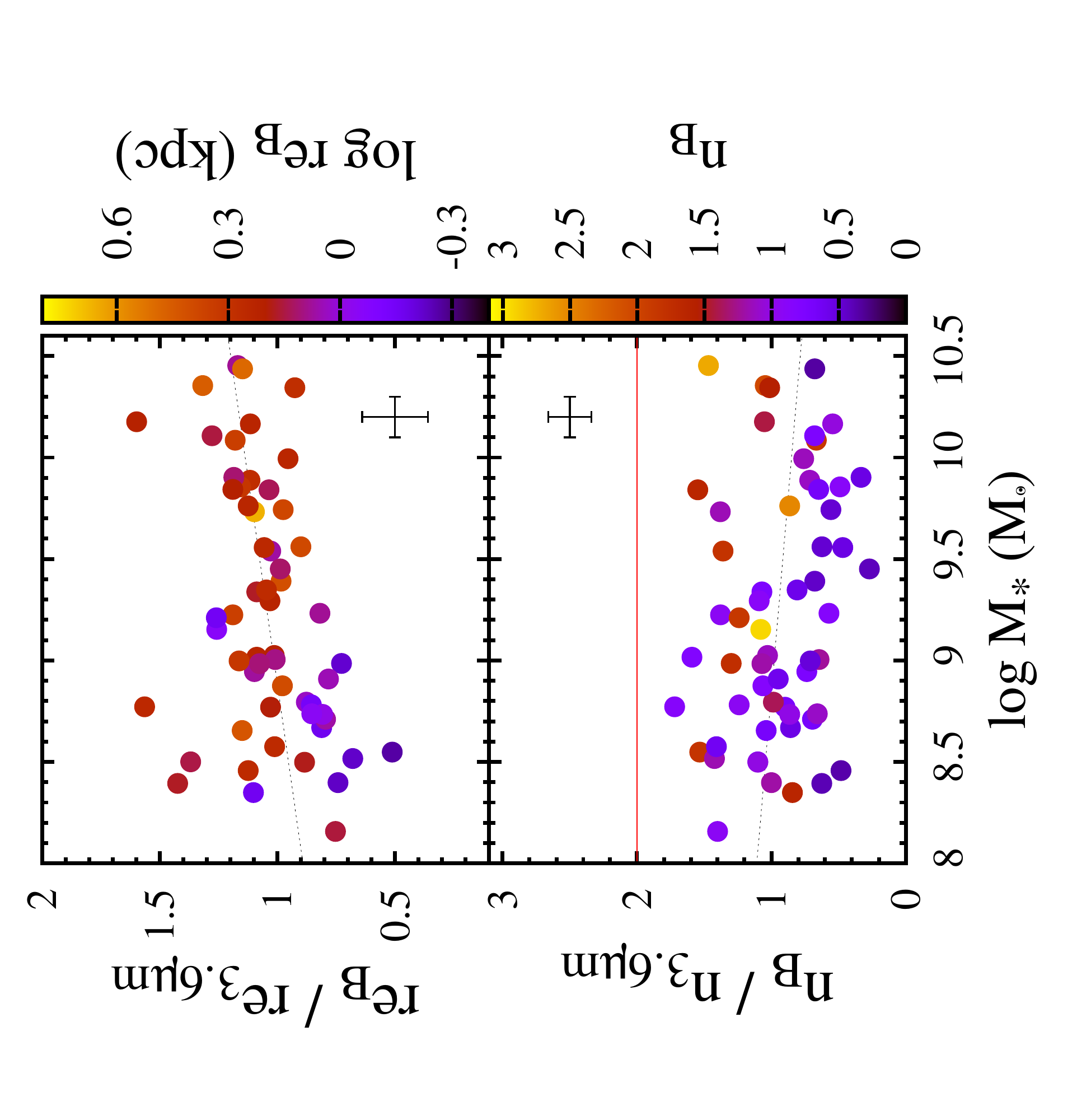} 
 \caption{The ratios of the half-light radii $\rm{r_{e}}$ (top) and S\'{e}rsic indices $n$ (bottom) of the B-band and 3.6$\rm{\mu m}$ against stellar mass $\rm{M_{*}}$ and colour 
 coded as a function of $\rm{(r_{e})_{B}}$ and $\rm{n_{B}}$. Typical error bars are shown in black in the corner of each plot. (A colour version of this figure is available in the online journal.)}
 \label{figure:n_re}
 \end{figure}

Before comparing these results to the literature data, we looked at the ratio of the $\rm{(r_{e})_{B}}$ to $\rm{(r_{e})_{3.6}}$ and the ratio of the
 $\rm{n_{B}}$ to $\rm{n_{3.6}}$ of the VGS galaxies in order to understand the galaxy concentration properties. In Figure~\ref{figure:n_re}, we present the $\rm{(r_{e})_{B}/(r_{e})_{3.6}}$ (top panel) and the $\rm{n_{B}/n_{3.6}}$ ratios (lower panel) as a 
function of stellar mass, $\rm{(r_{e})_{B}}$ and $\rm{n_{B}}$. We find that 
the $\rm{(r_{e})_{B}/(r_{e})_{3.6}}$ increases as a function of both increasing stellar mass and increasing $\rm{re_{B}}$. This shows that light in the smaller galaxies is more concentrated in B than
in 3.6 $\rm{\mu}$m. 
This could be caused by the fact that star formation is more concentrated in the smaller galaxies, or distributed more in the outer parts in the larger objects. Also, extinction could 
contribute to this
relation. Larger galaxies generally contain more dust. Since this causes more extinction in the centre of the galaxy, and the extinction in the B band is more substantial, it causes the 
effective radius $\rm{(r_{e})_{B}}$ in the B band to be larger than the one at $\rm{(r_{e})_{3.6}}$. In reality, both effects probably play a role. The large scatter for the faint galaxies 
could be caused by the range of central star formation properties in these objects. In the absence of extinction and young stellar populations, the $\rm{r_{e}}$ ratios that we observe could 
also be caused by metallicity gradients. Since our galaxies are spirals with both extinction and young stellar populations, this last option does not seem very likely, although it will have a small 
contribution.

In the lower panel of Figure~\ref{figure:n_re} we see that the the ratio of the Sersic indices $\rm{n_{B}}$ to $\rm{n_{3.6}}$ is on the order of unity. There is a slight tendency for the ratio 
to decrease with increasing stellar mass. It is a clear reflection of the more substantial extinction in larger galaxies, manifesting itself in a 
radial surface brightness profile that is less peaked 
in the B band. For most of the small galaxies there is no difference, suggesting a smaller effect. It also means that the star 
formation histories do not vary drastically as a function of radius.

\begin{figure}
 \centering\includegraphics[scale=0.43]{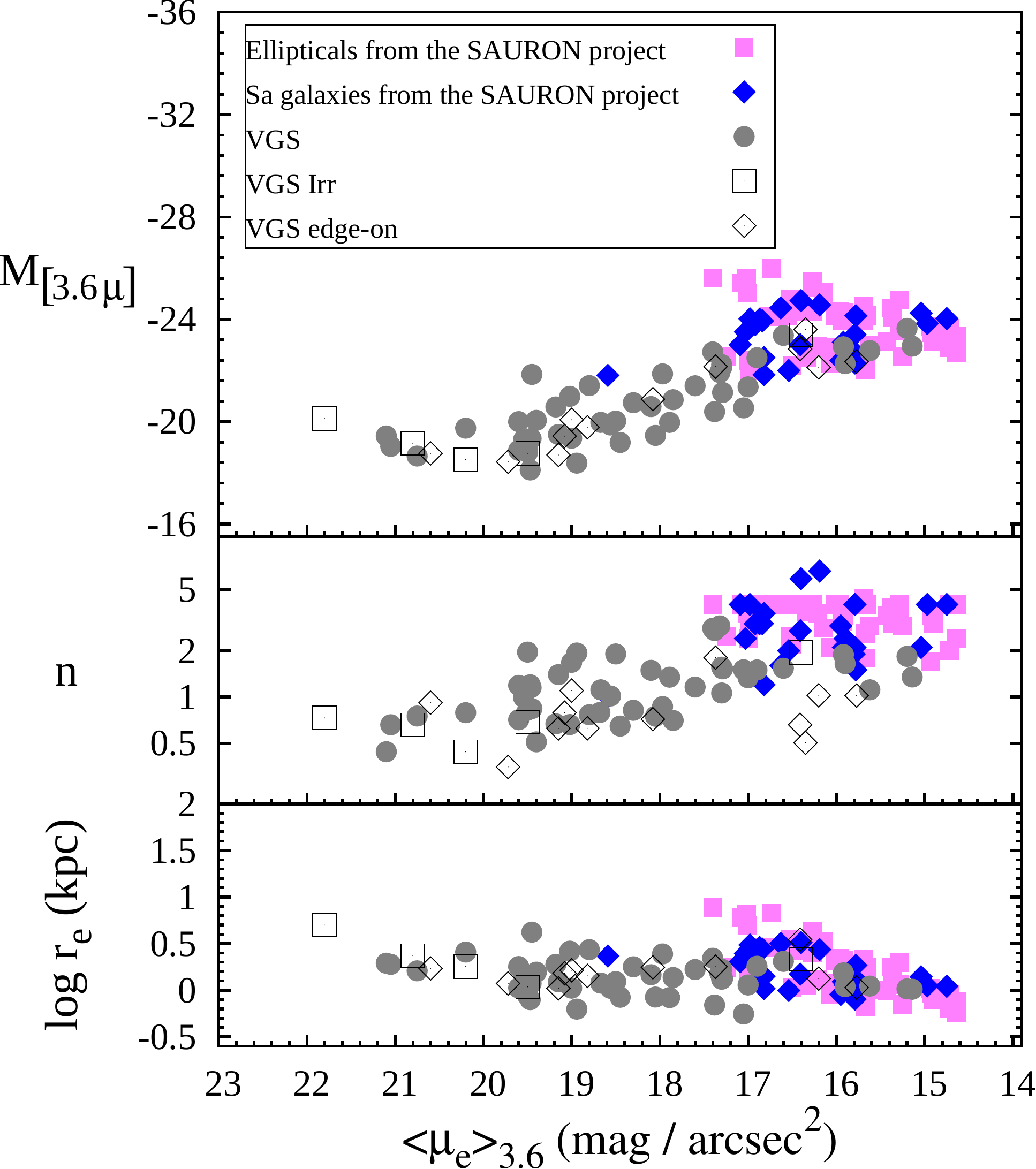} 
 \caption{Absolute magnitude, S\'{e}rsic index $n$ and the logarithm of the half-light radius $\rm{r_{e}}$ shown against the surface brightness at $\rm{r_{e}}$ $\rm{(\mu_{e})}$ for the Spitzer 
[3.6] band using a sample of galaxies that has been observed by the SAURON collaboration \citep{barroso2011}. Here 
 we use the mean effective surface brightness $\rm{\langle\mu_{e}\rangle_{3.6}}$ instead of the effective surface brightness. (A colour version of this figure is available in the online journal.)}
 \label{figure:sauron}
 \end{figure}

To get a first broad impression of how the near-infrared 3.6 $\rm{\mu m}$ structural parameters of the VGS galaxies compare to 
those of a more average population of galaxies, we compare these to some literature samples of ellipticals and spirals. In 
Figure~\ref{figure:sauron}, we compare the VGS galaxies with a sample of galaxies that has been observed with the Spectrographic Areal Unit for Research on Optical Nebulae 
(SAURON) and for which structural parameters have been published by \cite{barroso2011}. The VGS galaxies are considerably fainter than the spiral and elliptical galaxies. In this sample they are 
on average 2-3 magnitudes fainter. In the near-infrared regime, the VGS galaxies also have smaller S\'{e}rsic indices than 
ellipticals. On the other hand, VGS galaxies have near-infrared effective radii quite similar those of ellipticals and 
Sa galaxies. 

In order to compare the structural parameters of our sample to many  galaxies of different morphological types we use different structural parameters such as the central surface brightness $\rm{(\mu_{0})_{B}}$ and the scale length $h$
in addition to the surface brightness at $\rm{(r_{e})_{B}}$, the half-light radii $\rm{(r_{e})_{B}}$, S\'{e}rsic index $n$ and absolute magnitude $\rm{M_{B}}$. 
Our comparison samples have not been selected on the basis of environment. Instead, they are taken from a number of papers that present carefully determined photometry of a range of galaxies.

\begin{figure*}
  \centering
  \includegraphics[scale = 0.4]{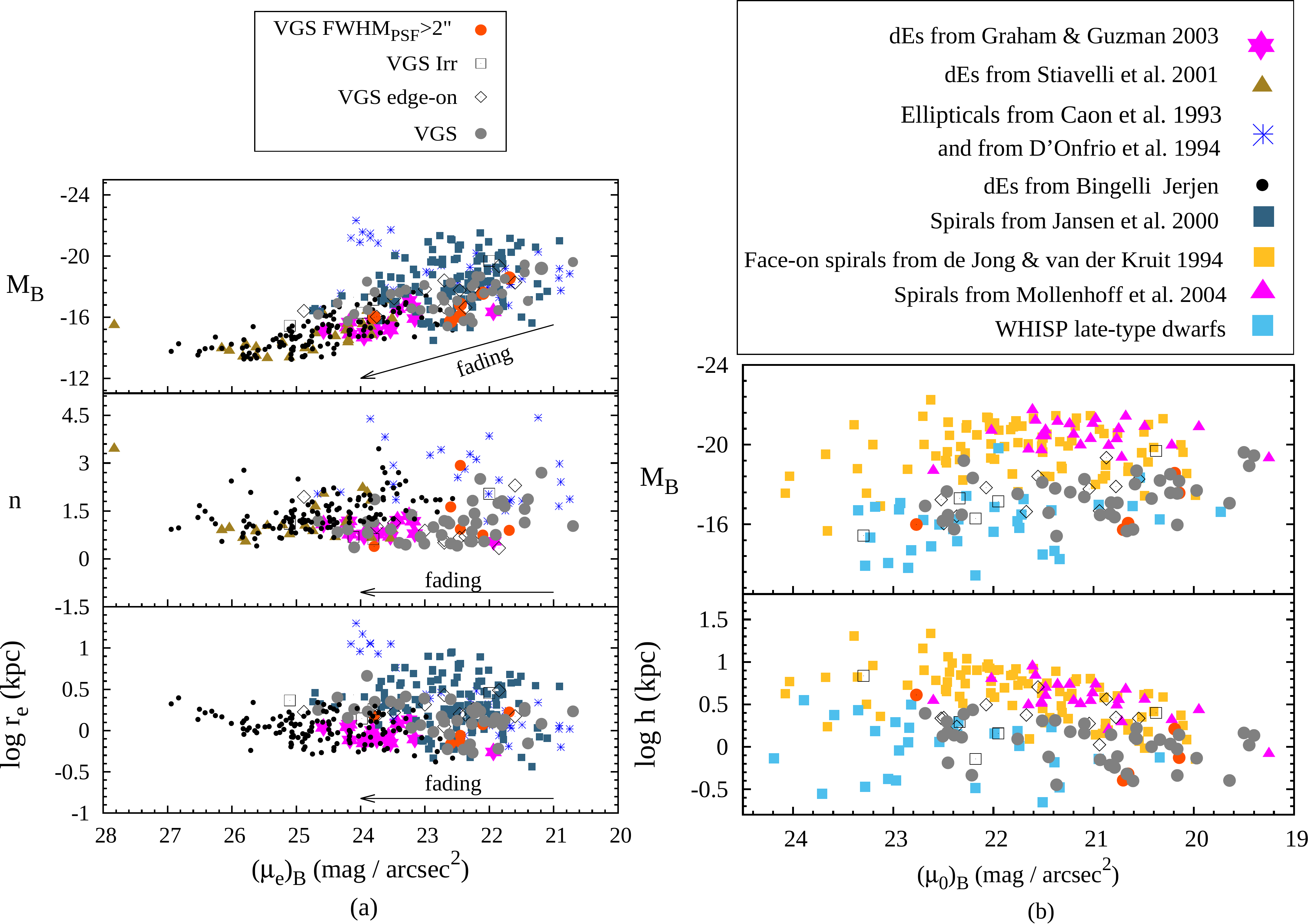}
\caption {Comparison of the structural parameters of the VGS galaxies to the galaxies of various morphologies. a) Absolute magnitude $\rm{M_{B}}$, S\'{e}rsic index $n$ and the logarithm of the half-light radius 
$\rm{r_{e}}$ shown against the surface brightness at $\rm{r_{e}}$ $\rm{(\mu_{e})}$ for B-band. VGS galaxies are compared to the dwarf ellipticals dE and ellipticals. b) Absolute magnitude and scale 
length $h$ shown against the central surface brightness $\rm{(\mu_{0})_{B}}$ where VGS galaxies are compared to spirals. In both panels, the same symbols in Figure~\ref{figure:phot}a are used to represent 
the VGS galaxies. Pink stars represent the Coma dEs from \citet{graham2003}, dots represent dE galaxies from \citet{bing1998}, green triangles represent dE galaxies from \citet{stia2001}, black 
asterisks represent intermediate to bright E galaxies from \citet{canon1993} and \citet{don1994}. Blue-green squares are spiral galaxies from \citet{jansen2000}. Yellow squares represent 
the face-on spiral galaxy sample from \citet{dejong1994} and pink triangles are the spiral galaxy sample from \citet{mol2004}. Finally light-blue squares are late-type dwarf galaxies of WHISP survey from 
\citet{swat2002}. (A colour version of this figure is available in the online journal.)} 
 \label{figure:phot_comp}
\end{figure*}

In Figure~\ref{figure:phot_comp}a we compare our B-band photometry results with the same data used in 
\cite{graham2003} (Figure 9, middle panel in that paper) and in \cite{jansen2000}. The data used in \cite{graham2003} consists of a variety of dE and elliptical galaxies. There are 18 dE galaxies 
from the Coma Cluster selected by \cite{graham2003}, a sample of 23 Virgo and Fornax Cluster dE galaxies from \cite{stia2001}, 
another sample of dE galaxies from the Virgo Cluster studied by \cite{bing1998}, and intermediate to bright elliptical galaxies 
from \cite{canon1993} and \cite{don1994}. The data adopted from \cite{jansen2000} consist of spirals of type Sb to Sm. They belong to a sample of 200 nearby galaxies selected 
from the first CfA redshift catalogue \citep{huchra1983}. All galaxies in this sample have been morphologically classified. It includes a wide range of late type galaxies over a large
range of luminosities ($\rm{M_{B}}$ = -14 to -22).

In Figure~\ref{figure:phot_comp}b, we use a sample of spiral and late-type dwarf galaxies from \cite{dejong1994}, 
\cite{mol2004} and \cite{swat2002}. The spiral galaxy sample of \cite{dejong1994} has been selected from the Upsala General 
Catalogue of Galaxies (UCG) in such a way that only the S and Sb type galaxies have been included. Spiral galaxies of \cite{mol2004} have 
been selected from the Revised Shapley-Ames Catalogue \citep{sand1981}. These include galaxies of Hubble type from Sa to Sc without 
a strong bar. The late-type dwarfs galaxies are a subsample of 171 late-type dwarf and irregular galaxies observed as part of the 
Westerbork $\rm{H\textsc{i}}$ Survey of Spiral and Irregular Galaxies (WHISP). Here we use 46 of them with B-band photometric data from \citet{swat2002}.

VGS galaxies appear to trace the same range of surface brightness, in terms of both $\rm{(\mu_{e})_{B}}$ and $\rm{(\mu_{0})_{B}}$, as spiral galaxies and late-type dwarfs 
(see Figure~\ref{figure:phot_comp}a and b). Within the same range of surface brightness, the scale lengths and absolute magnitudes of the VGS galaxies are also comparable to those of late-type
dwarfs in the WHISP sample and in the \cite{jansen2000} sample. In this context, we should note that the late-type dwarfs in the WHISP sample reach out to substantially lower surface brightness levels.
In comparison to more regular spiral galaxies, such as those in the spiral samples of the \cite{dejong1994} and \cite{mol2004}, VGS galaxies are substantially fainter and smaller. On the average, these
spirals are 2.5 magnitudes brighter than the VGS galaxies.

As may be seen from Figure~\ref{figure:phot_comp}a (top panel) the VGS galaxies are fainter than ellipticals. As expected, the ellipticals have a substantially larger S\'{e}rsic index $n$. In fact, the 
indices $n$ of ellipticals are larger than those of nearly all galaxies represented in Figure~\ref{figure:phot_comp}a (central panel). Interestingly, it turns out that the S\'{e}rsic indices as well as
the effective radii of dE galaxies are of the same order as those of the VGS galaxies. Nonetheless, the VGS galaxies are substantially brighter than the dE galaxies, both in terms of absolute magnitude 
and effective surface brightness. The mean absolute magnitude of $\sim$ $-$17.1 of the VGS galaxies is around $\sim$1 magnitude brighter than that of the dE galaxies. 

In summary, VGS galaxies have small half-light radii and scale lengths, rather similar to those of late-type dwarfs and small spirals. In terms of size and luminosity they therefore look very much like
late-type dwarfs. Interestingly, their S\'{e}rsic indices are similar to dEs indicating that they have similar light distributions. However, dE's have slightly smaller half-light radii, and a significantly 
smaller absolute brightness and effective surface brightness (Figure~\ref{figure:phot_comp}a).
\subsection{Environment and {H\,{\sc{i}}\,} content}
As explained in \cite{kreckel2011}, the void environments of the VGS galaxies are defined using a watershed delineation of the 
galaxy density field. The geometric and topological identification of the void regions, and the criterion that within the 
identified watershed basins the VGS galaxies should reside in regions with a galaxy density contrast $\rm{\delta}$ < $-$0.5 (averaged over 1 Mpc), 
ensures that they are located in the central interior of their voids. Within the sample, the VGS galaxies are found at an 
underdensity contrast $\rm{\delta}$ between $-$0.94 to $-$0.52, with an average underdensity value of $\langle \rm{\delta}\rangle 
\approx -0.78$ when 1 $h^{-1}$ Mpc Gaussian smoothing filter applied.

The void galaxy identification procedure for the VGS does not involve any criteria related to intrinsic galaxy 
properties like morphological type. At first glance, we would therefore not expect any dependence of the 
$\rm{H\textsc{i}}$ content and other characteristics of the VGS galaxies on the relative positions of the galaxies 
within their voids. However, we made some striking discoveries, in particular for several galaxies in the 
deepest realms of cosmic voids, i.e.. with $\rm{\delta}$ < $-$0.9. 

It is interesting that, of the five galaxies residing in desolate regions with $\rm{\delta}$ < $-$0.9, three are not detected 
in $\rm{H\textsc{i}}$. Two of these are galaxies with the smallest half-light-radius of $\sim$0.7 kpc in the VGS sample. They are the very compact galaxies 
VGS\_20 and VGS\_41 (Figure~\ref{fig:morph}). 
Perhaps the most surprising finding is that of an early-type galaxy, VGS\_05, which lies in the deepest underdensity of 
the entire sample, at $\rm{\delta \equiv -0.93}$.  VGS\_05 is indeed a rarity amongst the void galaxies. It has a stellar 
mass-to-light ratio of $\sim$6.8 in 3.6$\rm{\mu m}$, is not detected in $\rm{H\textsc{i}}$, and has no detected companion, neither in the optical nor 
in $\rm{H\textsc{i}}$. The presence of this galaxy in the sample is interesting, as it is the only example of this morphological type. 
Considering its high stellar mass-to-light ratio, it must have evolved much faster than the rest of the sample and consumed its 
$\rm{H\textsc{i}}$. The remaining two galaxies at $\rm{\delta}$ < $-$0.9 are more reminiscent of other void galaxies. They are 
detected in $\rm{H\textsc{i}}$ and are blue disk galaxies with half-light radii comparable to the rest of the VGS sample. 

When considering the position of void galaxies within the context of the overall structural complexity of the cosmic web, the 
two VGS galaxies with the most outstanding locations are VGS\_31 \citep{beygu2013} and VGS\_12 \citep{stanonik2009}. VGS\_31 is a system 
of three interacting system aligned along a filamentary feature in the interior of a void. It is likely that this system, 
discussed in detail in \cite{beygu2013}, reflects a density enhancement in an underlying tenuous dark matter filament. The 
other interesting case is that of VGS\_12, a polar disk galaxy \citep[for an extensive study see][]{stanonik2009}. It is located 
in a tenuous wall that forms the boundary between two walls, and is amongst the most isolated objects in the nearby Universe. Even 
though its colour, stellar mass and star formation activity are not dissimilar from other typical void galaxies, such as VGS\_20 and 
VGS\_41, it has the uniquely interesting feature of an immense $\rm{H\textsc{i}}$ disk rotating perpendicular to its stellar disk. 
This disk may hint at the influence of its special location, which may be responsible for the cold inflow of gas from the two 
voids on whose boundary it is located \citep{stanonik2009}. 

In terms of structural parameters, VGS\_12 has a half-light radius $\rm{(r_{e})_{B}\approx 1.2}$ kpc and a surface brightness 
$\rm{(\mu_{e})_{B}}$ $\approx$ 23 $\rm{mag/arcsec^{2}}$. These are comparable to the VGS sample in general. However, it has a significantly 
smaller S\'{e}rsic index $\textit{n}\approx 0.7$ than the mean $\textit{n}=1.1$ of the VGS sample. 
\begin{figure}
   \centering\includegraphics[scale = 0.34]{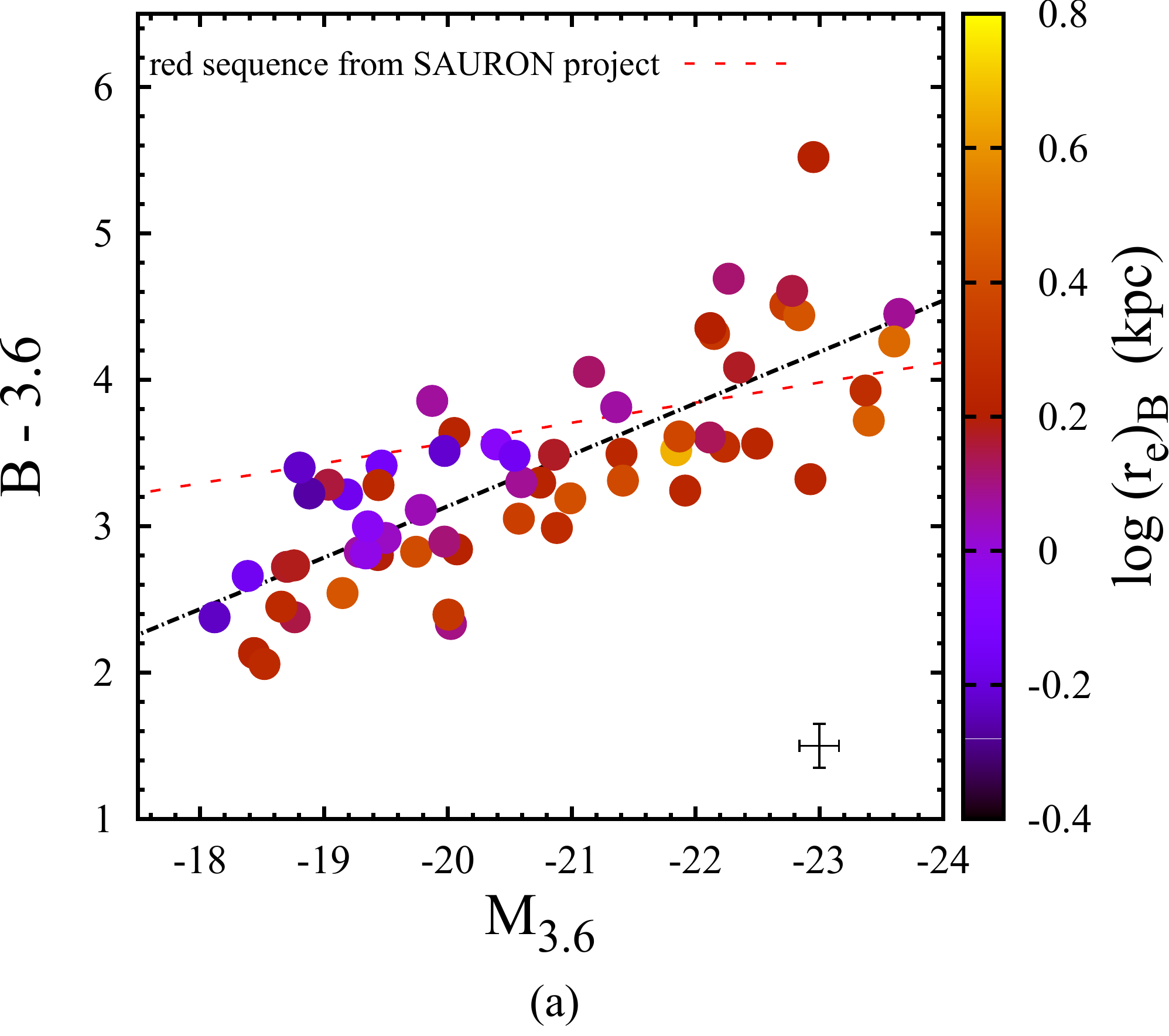}
   \centering\includegraphics[scale = 0.32]{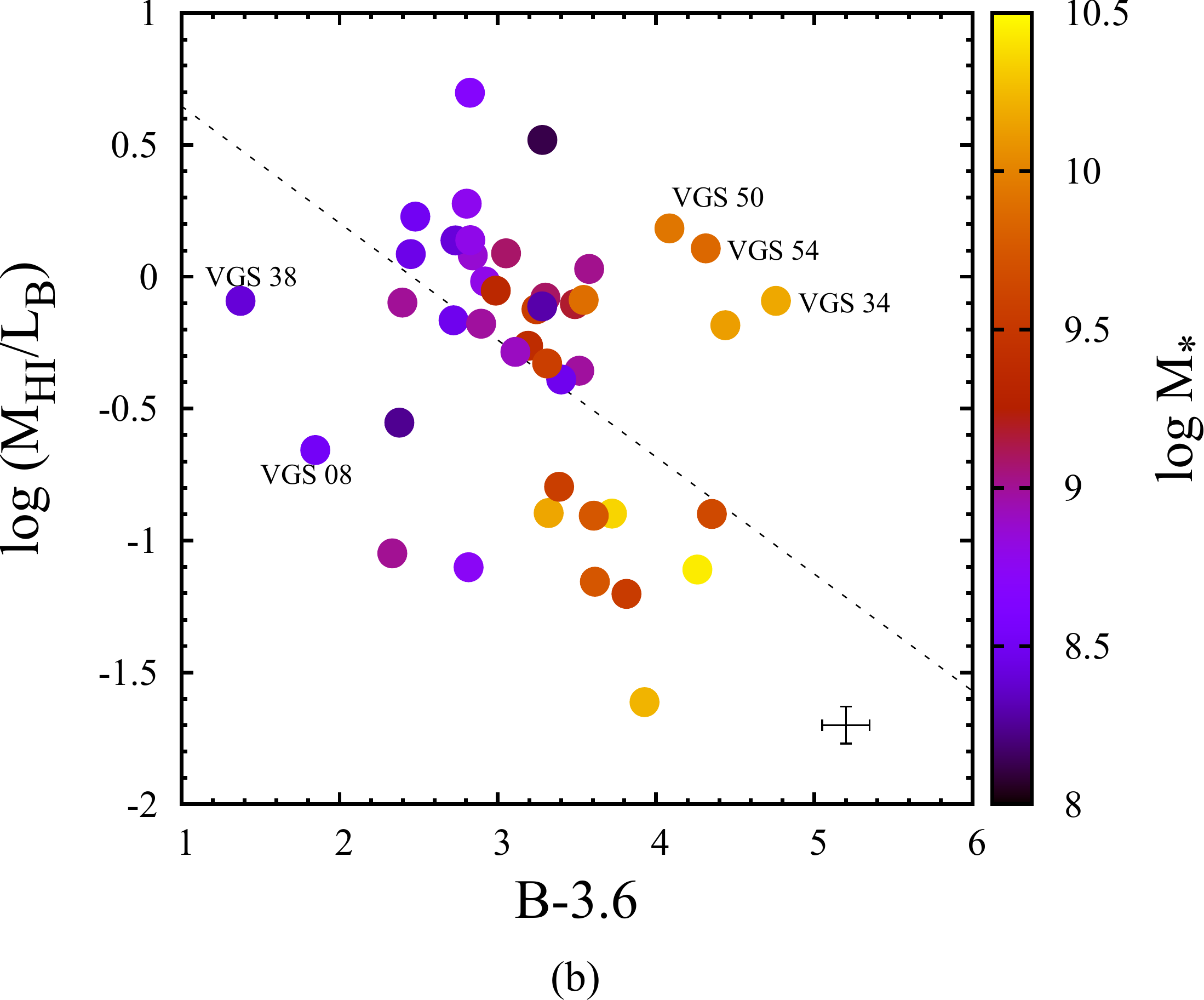}
\caption {Relation between the $\rm{B-[3.6]}$ colour, absolute magnitude $\rm{M_{3.6}}$, the $\rm{M_{H\textsc{i}}}$ mass-to-light $\rm{L_{B}}$ ratio, half light radius $\rm{(r_{e})_{B}}$ and the stellar mass $\rm{M_{*}}$.
Top: $\rm{B-[3.6]}$ colour (at half-light radii $\rm{r_{e}}$) against $\rm{M_{3.6}}$ and coloured with the logarithm of the $\rm{(r_{e})_{B}}$. With the red dotted line, we show the red sequence 
of \citet{barroso2011}. Bottom: The $\rm{M_{H\textsc{i}}}$ mass-to-light $\rm{L_{B}}$ ratio against the $\rm{B-[3.6]}$ colour as a function of the stellar mass $\rm{M_{*}}$. 
In both plots, $\rm{H\textsc{i}}$ non-detections are shown as upside-down triangles. Galaxies marked with their names are excluded from the linear regression due to their uncommonly high or 
low $\rm{H\textsc{i}}$ mass-to-light ratios (see text for details). (A colour version of this figure is available in the online journal.)} 
\label{figure:hlight-ratio}
\end{figure}
Two members of the VGS\_31 system 
have a surface brightness that is higher than the mean $\rm{(\mu)_{B}}$ $\approx$ 22.7 $\rm{mag/arcsec^{2}}$ of the VGS sample: for VGS\_31b 
$\rm{(\mu)_{B}}$ $\approx$ 20.7 $\rm{mag/arcsec^{2}}$ and VGS\_31a has $\rm{(\mu)_{B}}$ $\approx$ 22 $\rm{mag/arcsec^{2}}$. Both are also 
considerably larger than the average VGS galaxy, with their half-light radii in excess of the mean, $\rm{(r_{e})_{B}\approx 1.6}$ kpc 
of the VGS sample. 
\section{Stellar population evolution and star formation history}
\label{sec:stel_pop}
With the aim of assessing how the VGS galaxies have evolved, we have explored a scenario in which they remain in 
isolation, and gradually and uninterruptedly evolve as their current stellar population is ageing. In this scenario, 
galaxies will slowly become fainter and redder as their stellar populations become more and more dominated by older, 
fainter and redder stars. To investigate this evolutionary trend, we partially follow the work of \cite{meidt2014}.
In this study, stellar populations presented in the 3.6 $\rm{\mu}$m are purely based on stellar evolutionary 
tracks and observed stellar fluxes \citep{benjamin2003,church2009}. In all other stellar population models 
that present predictions for the 3.6 $\rm{\mu}$m band \citep{marigo2008,bruchar2003}, theoretical stellar atmospheres were
used. However, these can strongly over- or under-estimate the fluxes \citep{peletier2012}. 

Our calculations are based on the - strong - assumption that the stellar populations of each VGS galaxy have the 
same age and metallicity. Once the latter is chosen, we subsequently determine the age and stellar mass-to-light ratio in 
the B-band using the MILES models of \cite{vazdekis2010}. As a result the effects of fading, i.e. the ageing of a stellar population with time, a galaxy typically 
becomes fainter by $\sim$ 2.5 magnitude over 11.58 Gyr, implying that the mass-to-light ratios will increase by these numbers. The fading process 
can be investigated by studying its effect in scaling relations in B-band such as presented by the arrows in Figure~\ref{figure:phot_comp}.

\subsubsection{Colours, magnitudes and gas content}
During their evolution, galaxies are expected to consume their gas and form fewer stars as time goes by. As a result, they turn redder and their 
stellar mass-to-light ratios increase. Within this scenario, we expect that the VGS galaxies that have higher $\rm{H\textsc{i}}$ mass-to-light ratio have 
lower stellar mass-to-light ratio and vice versa. 

\begin{figure}
 \centering\includegraphics[scale=0.23]{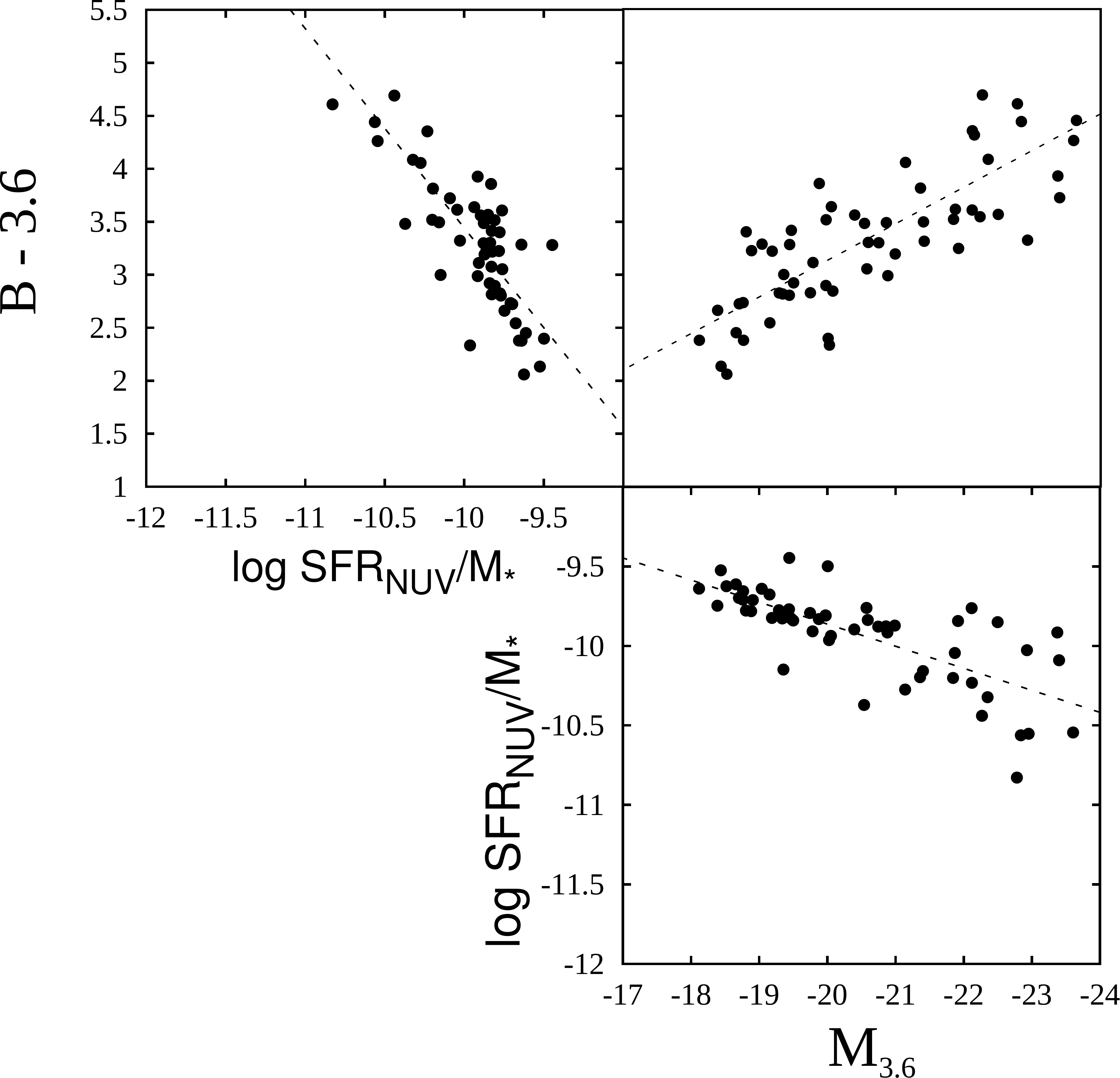} 
 \caption{Relations between the $B-[3.6]$ colour, specific star formation rate in the near-UV ($\rm{SFR_{NUV}/M_{*}}$) and the magnitude $\rm{M_{3.6}}$ are presented in three panels.  }
 \label{figure:col_sfr_M}
 \end{figure}

To see how viable this trend is for the VGS galaxies, we will use the results presented in Figures~\ref{figure:hlight-ratio} and 
~\ref{figure:col_sfr_M}. We have added the red sequence to this diagram, following the relation inferred by \cite{barroso2011} that concerns early-type galaxies observed with SAURON.
This sequence has been defined observationally by fitting a number of non-rotating elliptical galaxies, and assuming that these 
galaxies consist purely of old stars.

The first observation from top panel of Figure~\ref{figure:hlight-ratio}a is that the redder, and brighter, VGS galaxies have larger effective 
radii. The vertical distance to the red sequence indicates the blueing by young stellar populations. This is a measure of the 
galaxy age, independent of metallicity. We see that most of the massive VGS galaxies fall above the red sequence, indicating that 
these objects are reddened by dust. It also means that the $B-[3.6]$ colour cannot give much information about the stellar 
population of the galaxies. The smaller galaxies all lie below the red sequence. The distance to the red sequence increases with 
decreasing magnitude for the most massive galaxies. This indicates that the faintest galaxies are probably the youngest, and are beset by the lowest extinction.
The bottom panel of Figure~\ref{figure:hlight-ratio} shows the relation between $\rm{H\textsc{i}}$ mass-to-light ratio and $B-[3.6]$ colour as a function 
of stellar mass from the 3.6$\rm{\mu}$m photometry. The linear regression does not include the $\rm{H\textsc{i}}$ non-detections 
(marked by upside-down triangles), nor the galaxies marked with their names. As expected, and in accordance with \cite{kreckel2012}, 
the  $\rm{M_{HI}}$-to-$\rm{L_{B}}$ ratio is a decreasing function of stellar mass. 
There are a few galaxies with uncommonly high or low $\rm{M_{HI}}$-to-$\rm{L_{B}}$ ratios for their stellar mass and colour. In the 
diagrams, these are indicated by their names. VGS\_34, 50 and 54 are still $\rm{M_{HI}}$-rich, given the fact that they are red and do not
actively form stars (upper left panel of Figure~\ref{figure:col_sfr_M}). On the other hand, VGS\_08 and VGS\_38 have much lower 
$\rm{M_{HI}}$-to-$\rm{L_{B}}$ ratios than should be expected, given the fact that they are quite blue and are star-forming galaxies 
(Figure~\ref{figure:col_sfr_M}). These galaxies have an irregular shape and the $\rm{H\textsc{i}}$ disk
 of VGS\_38 shows in particular strong signs of interaction and therefore has large uncertainties in the estimate of the total 
$\rm{H\textsc{i}}$ mass. This is also true for VGS\_25 and VGS\_31b.
\subsubsection{Star formation activity and history}
Figure~\ref{figure:col_sfr_M} shows the relations between the $B-[3.6]$ colour, specific star formation rate in the near-UV ($\rm{SFR_{NUV}/M_{*}}$) and the absolute magnitude $\rm{M_{3.6}}$. 
From top to bottom, we first present the $B-[3.6]$ colour as a function of the $\rm{SFR_{NUV}/M_{*}}$ and as a function of the $\rm{M_{3.6}}$, then the $\rm{SFR_{NUV}/M_{*}}$
as a function of the $\rm{M_{3.6}}$. VGS\_05 is excluded from the plots. 

The Near-UV star formation rates ($\rm{SFR_{NUV}}$) have been calculated 
from the Galaxy Evolution Explore (GALEX) near-UV luminosities, corrected for internal dust attenuation
following the method outlined in \cite{schiminovich2010}.

\begin{equation}
 SFR_{UV} [M_{\odot} yr^{-1}] =\frac{L_{UV}f_{UV}(young)10^{0.4A_{UV}}}{\eta_{UV}},
\end{equation}
where $\rm{L_{UV}}$ is the luminosity in $\rm{erg\; s^{-1} Hz^{-1}}$, $\rm{f_{UV} (young)}$ is the fraction of light that originates
in young stellar populations, $\rm{\eta_{UV}}$ is the conversion factor between UV luminosity and recent-past-
averaged star formation rate and $\rm{A_{UV}}$ is the dust attenuation. Following \cite{schiminovich2010}, we assume $f_{UV} (young)$ = 1 and $\rm{\eta_{UV}}$ = $\rm{10^{28.165}}$. 
\citep[see][for details]{beygu2016}.

We find, not unexpectedly, that the $B-[3.6]$ colour is an increasing function of the specific star formation measured in the near-UV:
\begin{equation}
B-[3.6]= -1.88 \times \rm{SFR_{NUV}/M_{*}}-15.37,
\end{equation}
(top-left panel Figure~\ref{figure:col_sfr_M}) and of the absolute magnitude $\rm{M_{3.6}}$,
\begin{equation}
B-[3.6]= -0.34 \times \rm{M_{3.6}}-3.76,
\end{equation}
(top-right panel Figure~\ref{figure:col_sfr_M}). 

The specific star formation is a linear function of the absolute magnitude $\rm{M_{3.6}}$ as well. However, the slope of 0.14 is much smaller 
than the gradient of the ($B-[3.6]$) versus $\rm{SFR_{NUV}/M_{*}}$ relation and the ($B-[3.6]$) versus $\rm{M_{3.6}}$ relation. The logarithm of the 
$\rm{SFR_{NUV}/M_{*}}$ changes only $\sim$ 1 dex over a $\sim$ 5.5 magnitude range in absolute magnitude $\rm{M_{3.6}}$. This suggests that both
the near-UV and $B-[3.6]$ are probing the star formation properties of the VGS galaxies. More detailed 
modelling is needed to find out how these two observables can be combined. This is not straightforward, given the effects of extinction.
\section{Discussion}
\label{sec:discussion}
Given the often isolated environment, 
we may expect that the majority of void galaxies evolve secularly. The effect of the void environment is mainly that of limiting 
the mass of the galaxies that emerge in their interior. This is simply because of the absence of a sufficient amount of matter: the 
galaxy (halo) mass function is significantly shifted to lower masses \citep[see e.g][for a systematic appraisal of this 
effect]{aragon2010b,marius2014}. However, this raises the question of why the equivalent low-mass galaxies in higher density 
regions are not more seriously affected by environmental physical processes. The answer may lie in the past. At the time of their 
formation, void environments also involved a considerably higher density environment than we see at present. This can indeed be 
seen in computer simulations of structure formation \cite [see e.g.][]{springmillen2005,vogel2014}. 

This should lead us to investigate in more detail the possibly direct environmental impact on void galaxies, which in most situations will have been restricted to earlier cosmic epochs. However, 
in some situations there may be a hint of more direct environmental impact on a galaxy's life, or of a less straightforward evolutionary path, in the population 
of void galaxies that we have observed in the VGS survey. 

At this point the VGS and some of the recent studies draw our attention to the following findings. First, void galaxies
with stellar masses $\leq$ $\rm{10^{10}}$ $\rm{M_{\odot}}$ are blue late type star forming galaxies. Their star formation properties are not significantly different from the galaxies that reside in fields,
or moderate density environments \citep{alp2015,penny2015,beygu2016} suggesting that stellar mass is the dominant factor which regulates the galaxy properties. They are gas rich galaxies with diverse
$\rm{H\textsc{i}}$ properties, from regular rotation to disturbed gas morphologies and kinematics \citep{kreckel2011,kreckel2012}. The $\rm{H\textsc{i}}$ irregularities are evidence of 
ongoing interactions and gas accretion while regular rotating disks suggest secular evolution \citep[see][for details]{kreckel2012}.

Second, if the effect of the void environment is mainly that of limiting the mass of the void galaxies, than is there an upper limit on the stellar mass? The void galaxy sample of the VGS doesn't 
include any galaxy with stellar mass $>$ 3 $\times$ $\rm{10^{10}}$ $\rm{M_{\odot}}$ suggesting that the void environment has an effect on the size of galaxies, whereas the void galaxy 
sample of \cite{alp2015} and \cite{penny2015} have massive galaxies with stellar mass up to 5 $\times$ $\rm{10^{11}}$ $\rm{M_{\odot}}$. Moreover, these massive void galaxies regardless of their star formation 
activity seem to be disk galaxies. Based on their colour 
and morphology \cite{penny2015} argue that these massive void galaxies could have extinguished their gas supply, and have not undergone the interactions and mergers required to 
transform them into elliptical galaxies due to their isolation. On the other hand one of the three galaxies (VGS$\_$05) in the VGS with stellar mass of the order 
of $\rm{10^{10}}$ $\rm{M_{\odot}}$, is an elliptical galaxy with no star formation activity detected either in optical or near-infrared and not detected 
in $\rm{H\textsc{i}}$. The other two VGS galaxies are star forming spirals.

This raises the question how do void galaxies acquire their mass and can voids harbour galaxies as massive as those found in dense environments? Massive ellipticals are believed 
to accumulate mass via interactions and major mergers whereas disk galaxies via star formation and accretion of small satellites \citep{sch2000,kauffmann2003,kauffmann2004,dok2005}. 
Given its stellar mass 
of $~$2$\times$ $\rm{10^{10}}$ $\rm{M_{\odot}}$, VGS$\_$05 is not a typical massive elliptical galaxy but an early-type galaxy that reside very close to the centre of 
its parent void ($\rm{\delta \equiv -0.93}$). Is VGS$\_$05 an example of secular evolution or was it subject to interactions and major mergers very early in time compared to the 
rest of 58 void galaxies 
in the VGS sample? The same is true for the second early-type galaxy in the VGS, VGS\_24, that is an AGN.

On the other hand the existence of pairs and group of galaxies living in voids which show strong signs of interactions presents evidence for non secular galaxy evolution. Two such examples 
are VGS\_31 and VGS\_38. VGS\_31 is a group of three interacting galaxies embedded in the same $\rm{H\textsc{i}}$ envelope \citep[see][for details]{beygu2013}. The group also shows sign 
for minor merger and 
triggered star formation activity. We explored the dynamical and environmental origin and development of 
VGS\_31 in some detail \citep{rieder2013}. Our study revealed the often surprisingly complex and diverse dynamical 
evolution of the intravoid cosmic web of tenuous --- usually underdense --- filamentary and sheetlike structures. Quite reminiscent 
of the possibility of such an imprint of large-scale evolution on the nature of a void galaxy is the case of the polar ring 
galaxy VGS\_12 \citep{stanonik2009}. VGS\_38 in a group of three galaxies has clearly undergone major merger as seen in the 
$\rm{H\textsc{i}}$ velocity map \citep{kreckel2011,kreckel2012} and optical morphology (Figure 5), is another example for rare major merger events in voids. 

Galaxies grow via the inflow and infall of gas from the environment. From numerous simulation studies, we are beginning to  
understand that the morphology of galaxies may be determined to a considerable extent by the nature of the web-like 
structures in which they are embedded. The gas is channelled out of voids and transported via gaseous tendrils into the 
emerging galaxies \citep[e.g.][and references therein]{dekel2009,hahn2010,pichon2011,codis2012,rieder2013,aragon2013,miguel2016}. The void galaxies in the VGS survey, located in 
a well-defined large-scale environment, will provide us with crucial information on how their structure and morphology, 
as well as their star formation activity, may indeed be dependent on not only mass but also on the environmental 
dynamics indicated by the most recent cosmological simulations. 

\section{Conclusion}
\label{sec:conclusion}
We analysed Spitzer 3.6 and 4.5$\rm{\mu m}$ and B-band imaging of 59 void galaxies as part of the Void Galaxy Survey (VGS) to study their colour, stellar mass, 
galaxy concentration, morphology and star formation properties. The main conclusions are:
\begin{itemize}

\item We find that our void galaxy sample mostly consists of late-type galaxies. Most of them are similar 
to (Sd-Sm) galaxies, although a few are earlier type spirals and some are irregularly shaped galaxies.

\item The VGS galaxies have small half-light radii and scale lengths, rather similar to those of late-type dwarfs and small spirals. In terms of size, morphology and colour properties they clearly 
resemble late-type galaxies.

\item Interestingly, the light distributions of VGS galaxies bear some resemblance to that of dE galaxies. Like the latter, in both wavelength regimes their S\'{e}rsic indices are smaller than 2, $n$ $<$ 2.

\item They span a wide colour range in $B-[3.6]$ and most of them are blue, gas rich and star forming 
galaxies. An occasional VGS galaxy is gas poor, small and blue. However, they cover a considerable range of morphological types 
and different star formation properties \citep{beygu2016}. The same is true for their hydrogen gas content \citep{kreckel2012}. 

\item The voids in our sample do not appear to be populated by a particular type of void galaxy and despite the very low-surface-brightness limit of our B-band images, we have not found any 
dwarfs or small galaxies with $\rm{M_{*}}$ $<$ $10^{7}$ $\rm{M_{\odot}}$. After deriving the stellar masses using near-infrared images, we confirmed the upper 
limit of $10^{10.5}$ $\rm{M_{\odot}}$ inferred by \cite{kreckel2012}. We can conclude that the most prominent effect of the void environments is that it prohibits the formation of large and 
massive galaxies.
\end{itemize}

\section*{Acknowledgements}
The authors wish to thank the anonymous referee, for a careful assessment of the manuscript and very useful comments. We would like to thank M. Querejeta and S. Meidt for sharing their S4G nearby galaxy stellar mass catalogue.
This work was supported in part by the National Science Foundation under grant no. 1009476 to Columbia University. We are grateful 
for support from J. H. van Gorkom's Da Vinci Professorship at the Kapteyn Astronomical Institute. 
J.M. van der Hulst acknowledges support from the European Research Council under the European Union's Seventh Framework Programme (FP/2007-2013)/ ERC Grant Agreement nr. 291531.
K. Kreckel acknowledges grant KR 4598/1-2 from the DFG Priority Program 1573.
The Isaac Newton Telescope is operated on the island of La Palma by the Isaac Newton Group 
in the Spanish Observatorio del Roque de los Muchachos of the Instituto de Astrof\'{\i}sica de Canarias. This work is based on observations made with the Spitzer Space Telescope, 
which is operated by the Jet Propulsion Laboratory, California Institute of Technology under a contract with NASA. Support for this work was provided by NASA.

\begin{table*}
\caption{Col.(1): Galaxy name. Col.(2): SDSS Petrosian half-radius in SDSS r-band. Col.(3): SDSS Petrosian half-radius in SDSS g-band. Col.(4): B-band S\'{e}rsic index. 
 Col.(5): Effective radius in B-band. Col.(6): Effective surface brightness in B-band. Col.(7): [3.6] S\'{e}rsic index. Col.(8): Effective radius in [3.6]. 
Col.(9): Effective surface brightness in [3.6]. 
 Col.(10): Absolute magnitude in B-band. Col.(11): Absolute magnitude in [3.6]. 
 Col.(12): $B - [3.6]$ colour. Col.(13): Seeing. (The full table is available online).}
 \label{table:2}
 \begin{tabular}{@{}c@{\:}c@{\:}c@{\:}c@{\:}c@{\:}c@{\:}c@{\:}c@{\:}c@{\:}c@{\:}c@{\:}c@{\:}c@{\:}}
\hline
Galaxy \space & $\rm{(r_{P50})_{r}}$ \space &  $\rm{(r_{P50})_{g}}$ \space & $\rm{n_{B}}$ \space & $\rm{(r_{e})_{B}}$ \space & $\rm{(\mu_{e})_{B}}$ \space & $\rm{n_{3.6}}$ \space & $\rm{(r_{e})_{3.6}}$ \space & $\rm{(\mu_{e})_{3.6}}$ \space & $M_{B}$ \space &
        $M_{3.6}$ \space & (B -3.6) \space & $FWHM_{PSF}$ \\
 & $(kpc)$ & $(kpc)$ & & $(kpc)$ & ($mag/arcsec^{2}$) &  & $(kpc)$ & ($mag/arcsec^{2}$) & ($mag$) & ($mag$) & & ($arcsec$) \\
        (1)& (2) & (3) & (4) & (5) & (6) & (7) & (8) & (9) & (10) & (11) &(12) &(13) \\
\hline
  VGS\_01 & 0.88 & 0.8 & 0.75 & 1.18 & 22.29 & 1.02 & 1.05 & 19.47 & -19.87 & -17.57 & 2.3&3\\
  VGS\_02 & 1.63 & 1.73 & 0.81 & 1.8 & 23.96 & 0.51 & 1.57 & 20.35 & -20.05 & -16.48 & 3.58&1\\
  VGS\_03 & 0.76 & 0.76 & 0.55 & 0.87 & 22.21 & 0.65 & 0.83 & 19.15 & -19.19 & -15.97 & 3.22&1.3\\
  VGS\_04 & 0.55 & 0.53 & 2.93 & 0.87 & 21.13 & 2.71 & 0.7 & 17.58 & -20.39 & -16.83 & 3.56&2.3\\
  VGS\_05 & 2.1 & 2.17 & 1.86 & 2.9 & 23.42 & 2.8 & 2.2 & 18.57 & -22.73 & -17.64 & 5.08&1.6\\
 \hline 
\end{tabular}
\end{table*}

\bibliographystyle{mnras}
\bibliography{beygu} 

\bsp	
\label{lastpage}

 \end{document}